\documentstyle [floats,preprint,epsf,aps,eqsecnum]{revtex}

\global\arraycolsep=2pt         % Fix space around '=' in eqnarray's

\makeatletter
\def\footnotesize{\@setsize\footnotesize{9.5pt}\xpt\@xpt
\abovedisplayskip 10pt plus2pt minus 5pt
\belowdisplayskip \abovedisplayskip
\abovedisplayshortskip \z@ plus 3pt
\belowdisplayshortskip 6pt plus 2pt minus 2pt
\def\@listi{\topsep 6pt plus 2pt minus 2pt
\parsep 3pt plus 2pt minus 1pt \itemsep \parsep}}
\makeatother

\begin{document}

\title{Nuclear Reaction Rates in a Plasma}

\author{Lowell S. Brown}

\address{
Department of Physics, University of Washington
\\
Seattle, Washington 98195
\\}

\author{R. F. Sawyer}

\address{
Department of Physics, 
University of California at Santa Barbara
\\
Santa Barbara, California 93106
\\}

\maketitle

\newpage

\begin{abstract}

The problem of determining the effects of the surrounding plasma on
nuclear reaction rates in stars is formulated {\it ab initio}, using
the techniques of quantum statistical mechanics. Subject to the
condition that the nuclear reactions ensue only at very close approach
of the fusing ions and the condition that the reaction be slow, the
authors 
derive a result that expresses the complete effects of Coulomb barrier
penetration and of the influence of the surrounding plasma in terms of
matrix elements of well defined operators. The corrections do not
separate into the product of initial state and final state
effects. When the energy release in the reaction is much greater than
thermal energies, the corrections reduce, as expected, to evaluation
of the equilibrium probability of one ion being very near to the
position of another ion. We address the calculation of this
probability in an approach that is based on perturbation theory in the
couplings of the plasma particles to the two fusing particles, with
the Coulomb force between the fusing particles treated
nonperturbatively and interactions among the plasma particles treated
in the one-loop approximation. We recapture standard screening
effects, find a correction term that depends on the quantum mechanical
nature of the plasma, and put an upper bound on the magnitude of the
further correction terms for the case of a weakly coupled plasma. We
find that possible ``dynamical screening" effects that have been
discussed in the literature are absent. The form of our results
suggests that an approach that relies on numerical calculations of the
correlation functions in a classical Coulomb gas, followed by
construction of an effective two body potential and a quantum barrier
penetration calculation, will miss physics that is as important as the
physics that it includes.

\end{abstract}

\newpage

\section{Introduction and Summary}

The theoretical determination of fusion rates in stars [For example,
Clayton (1968)] divides into two steps: I. Finding the nuclear ``S
factor'' for specific reactions, which, loosely speaking,
characterizes what the rate of a specific nuclear reaction would be in
the absence of the Coulomb repulsion between the fusing ions. II.
Taking into account both the Coulomb repulsion between the fusing
particles and the influence of the surrounding charged plasma
[Reviewed in Ichimaru (1993)].  There is quite a clean division
between these two steps under conditions such that the energy release
in the fusion is very large compared with thermal energies, and such
that the reaction rate is small. Both of these conditions are well
satisfied in the case of nuclear reactions in the solar interior. In
such cases, the result that we seek in step II is, to first
approximation, the value at zero separation of the equilibrium
density-density correlation function of the fusing ions.\footnote{The
fusion reaction itself, and the associated nuclear forces, change the
behavior of wave functions at very short relative distances. But as
long as the reaction rate is very small, either because of the Coulomb
barrier penetration factor, or because the fusion interaction is weak,
these changes are significant only over a very small volume, and the
exterior plasma physics will have negligible effect on the space
variation of wave function within the volume. The quantities that
determine the influence of the plasma are then the correlation
functions that would exist with the nuclear charges concentrated at
points and the nuclear interactions turned off.}  We shall say nearly
nothing further about step I in this paper, and turn directly to the
barrier penetration and plasma aspect of the problem.

The history of this problem begins with the work of Gamow (1928), who
calculated the Coulomb barrier effects in the limit in which the only
Coulomb force taken into account is that between the two fusing
bodies, which gives the answer for the limit of an extremely dilute
plasma.  The first important corrections to the Gamow rates are found
in the results of Salpeter (1954) for the effects of Debye screening
by the electrons and ions in the plasma. These corrections, which will
serve as the benchmark for comparison with further corrections, are
given, at temperature\footnote{We measure temperature in energy units
so that the Boltzmann constant is unity, $k_B = 1$.} $T = \beta^{-1}$,
by
\begin{equation}
\Gamma_S = \Gamma_0 \exp\{ \beta e_1 e_2 \kappa_D \} \,,
\label{first}
\end{equation}
where $\Gamma_0$ is the rate in the absence of the plasma, $e_1 \,,
e_2$ are the charges of the fusing nuclei, and $\kappa_D$ is the
Debye wave number. This wave number squared is the sum of those for
each species in the plasma,
\begin{equation}
\kappa_D^2 = {\sum}_s \, \kappa_{D,s}^2 \,.
\end{equation}
Here we count each spin state of a particular kind of particle with a
separate species index $s$. 
The Debye wave number is generally defined by the rate of exponential
fall off of the charge-charge correlation function of the plasma. For
a dilute plasma, the one-loop or random phase approximation gives, as
is reviewed in Appendix A [Eq.~(\ref{onedebye})],  
\begin{equation}
\kappa_{D,s}^2 = 4\pi e_s^2 \, { \partial \langle n_s \rangle_\beta 
     \over \partial \mu_s } \,,
\label{qdebye}
\end{equation}
where $\langle n_s \rangle_\beta $ is the average density of species
$s$ which has charge $e_s = Z_s e$, and $\mu_s$ is the chemical
potential of this species. The general form may be
needed for the electron component in the plasma of a star in which the
effects of Fermi statistics may be important, but for the ionic
species, Boltzmann statistics suffices, in which case the Debye wave
number assumes the familiar form
\begin{equation}
\kappa_{D,s}^2 = 4 \pi e_s^2 \beta \langle n_s \rangle_\beta \,.
\label{cdebye}
\end{equation}

The effects of the plasma on reaction rates in the sun, as estimated
from the above formulae, are modest but important. For example, in the
entire region from the center of the sun to the radius inside which
one-half of the energy is generated, the screening effects give about
a 5\% increase in the rate of proton-proton fusion and a 20\% increase
in the rate for the reaction\footnote{These estimates use the tables
in Bahcall (1989), p.~90. The density, H/He ratio, and temperature all
change considerably over this region, but the screening factor almost
remains constant.} $ p + ^7{\rm Be} \to \gamma + ^8{\rm B}$.  In lower
temperature main sequence stars, and in highly evolved stars, the
corrections become even more important.

It is worthwhile exhibiting the physical origin of the Salpeter
correction shown in Eq.~(\ref{first}), the leading correction in the
dilute plasma case.  A plasma shields a charge brought into it, with a
point charge $q$ producing the Debye potential
\begin{equation}
\phi(r) = { q \over r} \, e^{ -\kappa_D \, r }  \,.
\label{debpot}
\end{equation}
Thus, assembling a charge distribution $\rho({\bf r})$ in a plasma has
an associated polarization energy --- the energy to assemble the
charge less the corresponding vacuum energy --- given by
\begin{equation}
\epsilon_\rho = {1 \over 2} \int\int (d{\bf r}) (d{\bf r}') 
\rho({\bf r}) \left[ { e^{ - \kappa_D | {\bf r} - {\bf r}' | }
 \over | {\bf r} - {\bf r}' | }  - {1 \over | {\bf r} - {\bf r}' | } \right]
\rho({\bf r}') \,.
\end{equation}
The limit for a point charge $q$ yields
\begin{equation}
\epsilon_q = - {1 \over 2} q^2 \kappa_D \,.
\label{epsilon}
\end{equation}
The Boltzmann factor associated with this polarization energy alters
the number density relation for the ions from that of a free gas to
read
\begin{equation}
\langle n \rangle_\beta = \lambda^{-3} e^{\beta \mu} 
e^{ - \beta \epsilon_q} \,,
\label{density}
\end{equation}
where $\lambda$ is the thermal wavelength defined by
\begin{equation}
\lambda = \sqrt{ {2\pi \beta \over M } } \,,
\label{donne}
\end{equation}
with $M$ the mass of the particle. A nuclear reaction rate depends
upon the probability to find two particles at the same spot, or more
precisely, upon the average of the product of the number densities of 
the two particles, $\langle n_1(0) n_2(0) \rangle$. With the two nuclear
particles having charges $e_1$ and $e_2$, this is given by
\begin{equation}
\langle n_1(0) n_2(0) \rangle_\beta = \lambda_1^{-3} \lambda_2^{-3}
e^{ \beta ( \mu_1 + \mu_2 ) } \exp\left\{ -\beta \, \epsilon_{(e_1 + e_2)} 
\right\}  \,.
\end{equation}
Thus, in view of the single-particle number density relation
(\ref{density}) and the form (\ref{epsilon}) of the polarization
energy, we find that
\begin{equation}
\langle n_1(0) n_2(0) \rangle_\beta = \langle n_1 \rangle_\beta 
\langle n_2 \rangle_\beta \exp \{ \beta e_1 e_2 \kappa_D \} \,.
\end{equation}
The exponential is the Salpeter correction in Eq.~(\ref{first}).

The method that we have just described is in the spirit of our work in
this paper which makes use of grand canonical ensembles that entail
chemical potentials. To make comparison with other treatments, we
should note that the same result is obtained if one replaces the
Coulomb potential between the fusing particles with a Debye screened
potential of the form of Eq.~(\ref{debpot}). Since only distances that
are short on the scale of the Debye length $ \kappa_D^{-1} $ enter
into the quantum tunneling, only the short-distance correction $\delta
V = - e_1 e_2 \kappa_D $ to the Coulomb potential $e_1 e_2 / r$ need
be retained. This is equivalent to an energy shift $ E \to E + e_1 e_2
\kappa_D $ or a momentum alteration $ \delta p = e_1 e_2 \kappa_D m /
p $. Thus the Gamow tunneling factor $ \exp\{ - 2\pi e_1 e_2 m / p \}
$ is corrected by the factor
\begin{equation} 
\exp\left\{
e_1 e_2 \kappa_D \left( { 2\pi e_1 e_2 m^2 \over p^3 } \right) 
\right\} \,.
\end{equation} 
In the thermal average, as we review below [{\em c.f.}
Eq.~(\ref{probp})], the relative momentum $p$ of the fusing particles
is replaced by the most probable momentum $ \bar p $, where
\begin{equation}
 \bar p^3 = 2\pi e_1 e_2 m^2 / \beta \,.
\end{equation}
This replacement produces the Salpeter correction in Eq.~(\ref{first}).

The ``weak screening" domain is defined as that in which the exponent
in Eq.~(\ref{first}) is small compared to unity.  This is also the
condition under which the Debye formula gives a good approximation to
the purely classical plasma correlations. Clearly, one way of going
beyond the Salpeter correction and toward the domain of the ``strongly
coupled plasma" is to use a correlation function that is better than
the one provided by the Debye formula. There is an extensive
literature devoted to the computation of the classical correlation
function for a plasma, and the determination of fusion rates therefrom
[Salpeter and Van Horne (1969); Dewitt, Graboske and Cooper (1973);
Graboske {\it et al.}\ (1973); Jancovici (1977)].  In these approaches one
translates a classical correlation function, usually numerically
determined, into a modification of the potential between the fusing
particles, and then calculates the quantum mechanical barrier
penetration factor, using this potential, in order to obtain the
fusion rate from a Gamow factor appropriate to this potential.  We
call this the ``basically classical" approach.

In this approach, there are conceptual problems raised by the division
of the problem into a quantum mechanical and a classical part. The
literature lacks any development that begins with a correct general
expression for the rate, shows how a division into classical
correlation plus quantum tunneling can be made as an approximation,
and gives a system for finding the corrections to this
approximation.\footnote{The work of Alastuey and Jancovici (1978) does
not completely meet these criteria for at least two technical reasons:
1) Dividing the system into subsystems of fusing particles and
surrounding plasma, the authors treat the plasma completely
classically, thereby missing the correction term that turns out to be
numerically most significant for the weakly coupled case. 2) Their
approximation scheme in which a numerically determined classical
correlation function for the plasma can be applied to the quantum
tunneling calculation for the fusing particles involves a zeroth order
in which the fusing particles are frozen in an ``average position" in
space. As a result the effects of center of mass motion are
irretrievably lost.}

We shall develop a general formulation of the problem which is based
on only one approximation --- that the fusing nuclei in the plasma can
themselves be described by Maxwell-Boltzmann statistics. In practice,
this is an excellent approximation as these particles are seldom so
dense as to require quantum statistics. No approximation is required
for the remainder of the plasma. This basic approximation enables us
to disentangle the internal Coulomb corrections in the fusion process
from the Coulomb interactions of the fusing nuclei with all the other
particles in the plasma (including other nuclei that could fuse). We
then employ this general formulation to obtain a straightforward
approach to the problem, based on perturbation theory. Considered as a
factor, the Salpeter correction differs from unity by a term of
leading order $e^3$ and therefore should be calculable in a
perturbation expansion. In describing our approach it will be
convenient to identify two particular ions as the ``fusing particles"
and all of the rest of the ions and electrons in the surrounding
medium as the ``plasma particles''.  As we shall see explicitly later,
the Salpeter correction can be derived as the leading correction from
a perturbative treatment in which the Coulomb force between the fusing
particles is treated exactly, the coupling between the fusing
particles and the plasma particles is treated perturbatively, and the
plasma itself is treated in one-loop order. The formulation does not
divide the development into a quantum mechanical and a classical part.

We shall develop this approach in some detail, and address the
corrections beyond the leading Salpeter term. We are still limited in
application to a weak screening domain of density and temperature, but
we are able to compare the correction terms with those that can be
produced in the basically classical approach. If we had found rough
agreement, then we could have argued that the basically classical
approach is a reasonable way to approach the regions of stronger
coupling. However we find the opposite, that the leading correction
terms that emerge from the complete perturbative approach have a
nature that is essentially different from those produced by the
basically classical approach. Our results thus point to the need for a
computational framework that does not divide into a classical and
quantum part.  

The new quantitative results applicable to the weak screening domain
are:

1) It is frequently stated that the Salpeter result, or static
screening, should be a good approximation only when the velocities of
the plasma particles are greater than the velocities of the fusing
particles, so that the latter see an average potential rather than a
time dependent one [e.g. Johnson {\it et al.}\ (1992), Shaviv \&
Shaviv (1996).]
This condition is rather badly violated for the ionic component of the
plasma.  But we find, in the leading order of $e^3$, no ``dynamical''
modifications of the Salpeter result in the ionic component of the
screening, in fact no dependence on the ionic masses at all.  This
conflicts with the results of Carrero, Sch\"afer, and Koonin (1988),
who found such terms. These ``dynamical" corrections to screening, if
present, would produce moderate decreases in predicted fusion rates
under solar conditions.  
In Appendix D we show how such terms, which never arise in our
approach, are produced in a real time treatment. We also find the terms
that cancel them.

2) There is one correction at the $e^4$ level that is numerically
significant under solar conditions.  The correction comes from a
quantum mechanical term in the electronic part of the plasma response
function, and is inaccessible to approaches based on a classical
treatment of the plasma.

3) A bound is put on all remaining corrections at the $e^4$ level.
This bound limits the contributions of such terms to around the 0.1\%
level, under solar conditions.

The sum of our new corrections turns out to be of marginal importance
to the various solar neutrino puzzles [Bahcall (1989), (1995)].
Thus we conclude, as have others, that the resolutions to these
puzzles will not be found in plasma physics.

\section{Reaction Rate Theory}

\subsection{General Framework}

A nuclear reaction, which we schematically indicate by
\begin{equation}
1 + 2 \to 3 + 4 \,,
\end{equation}
takes place over a very short distance in comparison with the particle
separations in a plasma. Hence, in the non-relativistic,
second-quantized formalism which we employ, it can be described by an
effective local Hamiltonian density
\begin{equation}
{\cal H}({\bf r},t) = {\cal K}({\bf r},t) + {\cal K}^\dagger({\bf r},t)
\,,
\end{equation}
in which
\begin{equation}
{\cal K}({\bf r},t) = g \, e^{-iQt} \, \psi^\dagger_4({\bf r},t)\,
\psi^\dagger_3({\bf r},t)\, \psi_2({\bf r},t) \, \psi_1({\bf r},t) \,,
\end{equation}
where $Q$ is the energy release of the reaction. The number density
operator of
a produced particle, say the particle labeled by 4, is given by
\begin{equation}
n(0) = \psi^\dagger_4(0) \psi_4(0) \,.
\end{equation}
Neglecting the kinematical particle flow described by the divergence
of the corresponding particle flux vector which vanishes in the
ensemble average, the time rate of change of this density is given by
\begin{eqnarray}
\dot n(0) &=& -i \int (d{\bf r}) \left[ n(0),{\cal H}({\bf r},0) \right]
\nonumber\\
         &=& -i \left\{ {\cal K}(0) - {\cal K}^\dagger(0) \right\} \,.
\end{eqnarray}
These operators create and destroy particles, and hence their
expectation vanishes in the unperturbed plasma ensemble which is
diagonal in particle number. The reaction rate $\Gamma$ appears in the
additional linear response to the action of the perturbation ${\cal
H}$,
\begin{eqnarray}
\Gamma &=& -i \int_{-\infty}^0 dt \int (d{\bf r}) \left\langle [ \dot n(0) ,
{\cal H}({\bf r},t) ] \right\rangle_\beta
\nonumber\\
       &=&  - \int_{-\infty}^0 dt \int (d{\bf r}) \left\langle \left[
{\cal K}(0) - {\cal K}^\dagger(0) , {\cal K}({\bf r},t) +
{\cal K}^\dagger({\bf r},t) \right] \right\rangle_\beta \,,
\end{eqnarray}
where $ \langle \cdots \rangle_\beta $ denotes the grand canonical
thermal average of the background plasma. 

Since the operator ${\cal K}$ does not conserve particle number,
\begin{equation}
\Big\langle \Big[ {\cal K}(0), {\cal K}({\bf r},t) \Big]
\Big\rangle_\beta = 0
= \Big\langle \Big[ {\cal K}^\dagger(0),{\cal K}^\dagger({\bf r},t)
\Big] \Big\rangle_\beta \,.
\end{equation}
Moreover, in virtue of the space-time translational invariance of the
ensemble and the antisymmetry of the commutator,
\begin{equation}
\left\langle \left[ {\cal K}^\dagger(0) , {\cal K}({\bf r},t) \right]
\right\rangle_\beta = - \left\langle \left[ {\cal K}(0) , 
{\cal K}^\dagger(-{\bf r}, -t) \right] \right\rangle_\beta \,.
\end{equation}
Thus, changing the sign of the integration variables in this term, 
$ ({\bf r},t) \to (-{\bf r}, -t) $, terms combine to produce an
integral over all times,
\begin{equation}
\Gamma = - \int_{-\infty}^{+\infty} dt \int (d{\bf r}) \left\langle
\left[ {\cal K}(0) , {\cal K}^\dagger({\bf r},t) \right] 
\right\rangle_\beta \,.
\end{equation}
The two terms which comprise the commutator which appear here
correspond to the destruction of particle 4 via $ 3+4 \to 1+2$ as well
as its creation by $ 1+2 \to 3+4$. We shall assume that initially
there are no particles of type 3 or 4 present in the plasma --- or at 
least that they are extremely dilute --- so that the operator 
${\cal K}^\dagger({\bf r},t)$ acting to the right on an initial state
vanishes. Hence, in this case, the only case that we shall consider,
\begin{eqnarray}
\Gamma &=& \int_{-\infty}^{+\infty} dt \int (d{\bf r}) \left\langle
 {\cal K}^\dagger({\bf r},t) {\cal K}(0) 
\right\rangle_\beta 
\nonumber\\
        &=& g^2 \int_{-\infty}^{+\infty} dt \int (d{\bf r}) \, e^{iQt} 
\left\langle \psi^\dagger_1({\bf r},t) \, \psi^\dagger_2({\bf r},t) \,
\psi_3({\bf r},t) \, \psi_4({\bf r},t) \, \psi^\dagger_4(0) \,
\psi^\dagger_3(0) \, \psi_2(0) \, \psi_1(0) 
\right\rangle_\beta \,.
\nonumber\\
&& \null
\label{genresult}
\end{eqnarray}

The result (\ref{genresult}) is equivalent to the `golden rule' rate
formula
\begin{equation}
\Gamma = {\sum}_{FI} P(I) | \langle F = 3,4;f | {\cal K}(0) | I = 1,2;i
\rangle |^2 (2\pi)^4 \delta ( E_F  - E_I - Q ) 
\delta ( {\bf P}_F  - {\bf P}_I ) \,.
\end{equation}
Here $ |I = 1,2;i \rangle$ describes the complete initial state of the
system --- the initial particles that take part in the reaction are
labeled by $1,2$ and the background plasma by $i$. The initial
particles $1,2$ are interacting with this background plasma, and so
their separate energy-momentum is not conserved. The initial states,
however, can be chosen to be eigenstates of the total energy-momentum
$E_I \,, {\bf P}_I$, and this we have done. The probability
distribution for the initial states is denoted by $P(I)$.  Similarly,
the final state $ \langle F = 3,4;f | $ contains the produced
particles $3,4$ and describes the final plasma $f$ with the total
energy-momentum of this state denoted by $E_F \,, {\bf P}_F$.  The
fusing particle coordinates and plasma coordinates are interlocked in
a complex way in the states, making the golden rule formula somewhat
ill adapted to proceeding further.  In applications in which reaction
energy release $Q$ is very large compared with thermal energies, or in
a case in which the final particles 3 and 4 are both neutral, it is
possible to express the rate in terms of a thermal correlation
function of particles 1,2 in the initial state, but we can proceed
quite some way from (\ref{genresult}) without making these
restrictions.

\subsection{Free Gas}

We pause now in our general development to illustrate the nature of the
result (\ref{genresult}) which we have just derived by considering the
limit in which the plasma is replaced by a free gas. For a free gas,
the ensemble expectation value in Eq.~(\ref{genresult})
factorizes. Since the particles 3 and 4 are produced out of the
vacuum, one factor is
\begin{equation}
\left\langle 0 \left| \psi_3({\bf r},t) \psi_4({\bf r},t) \psi^\dagger_4(0)
\psi^\dagger_3(0) \right| 0 \right\rangle = \int { (d{\bf p}_3) \over
(2\pi)^3 } \int { (d{\bf p}_4) \over (2\pi)^3 } \, e^{ i( {\bf p}_3 +
{\bf p}_4 ) \cdot {\bf r} - i (E_3 + E_4) t } \,,
\end{equation}
where
\begin{equation}
E_3 = { {\bf p}_3^2 \over 2 M_3 } \,, \qquad
E_4 = { {\bf p}_4^2 \over 2 M_4 } \,.
\end{equation}
The other factor is
\begin{equation}
\left\langle \psi^\dagger_1({\bf r},t) \psi^\dagger_2({\bf r},t)
\psi_2(0) \psi_1(0) \right\rangle_\beta = \int { (d{\bf p}_1) \over
(2\pi)^3 } n_1({\bf p}_1)
\int { (d{\bf p}_2) \over (2\pi)^3 } n_2({\bf p}_2)
e^{- i( {\bf p}_1 + {\bf p}_2 ) \cdot {\bf r} + i (E_1 + E_2) t } \,,
\end{equation}
where 
\begin{equation}
n_1({\bf p}_1) = {1 \over e^{\beta (E_1 - \mu_1)} \mp 1} 
\end{equation}
and
\begin{equation}
n_2({\bf p}_2) =  {1 \over e^{\beta (E_2 - \mu_2)} \mp 1}
\end{equation}
are the Bose or Fermi distributions of the initial particles with
chemical potentials $\mu_1$ and $\mu_2$ and energies
\begin{equation}
E_1 = { {\bf p}_1^2 \over 2 M_1 } \,, \qquad
E_2 = { {\bf p}_2^2 \over 2 M_2 } \,.
\end{equation}
With these results in hand, it easy to see that in the free gas limit,
the rate formula (\ref{genresult}) yields\footnote{Here we have
neglected possible spin variables. Spin degeneracy is readily
accounted for by including spin-weight factors in the density
distributions $n_1({\bf p}_1)$ and $n_2({\bf p}_2)$.}
\begin{eqnarray}
\Gamma_0 &=& g^2  \int { (d{\bf p}_4) \over
(2\pi)^3 } \int { (d{\bf p}_3) \over (2\pi)^3 } 
\int { (d{\bf p}_2) \over (2\pi)^3 } \, n_2({\bf p}_2)
\int { (d{\bf p}_1) \over (2\pi)^3 } \, n_1({\bf p}_1)
\nonumber\\
&& \qquad \qquad
(2\pi)^3 \delta ( {\bf p}_4 + {\bf p}_3 - {\bf p}_2 - {\bf p}_1 )
(2\pi) \delta ( E_4 + E_3 - E_2 - E_1 - Q) \,.
\end{eqnarray}

We use this free gas result to make several points. First of all, we
note that it reproduces the exact nuclear rate
\begin{equation}
\Gamma_0 = \int { (d{\bf p}_2) \over (2\pi)^3 } \, n_2({\bf p}_2)
\int { (d{\bf p}_1) \over (2\pi)^3 } \, n_1({\bf p}_1)
\int d\sigma v \,,
\end{equation}
if the coupling $g^2$ is taken inside the integration and replaced by
the squared nuclear matrix element $ |T|^2 $. In practice, the
effective Hamiltonian method produces the correct nuclear rate if the
coupling $g^2$ is replaced by an appropriate average value of the
squared matrix element $|T|^2$. Note that, as yet, the rate formula
does not account for the Coulomb interactions between the reacting
particles. Finally, we note that for most applications such as the
plasma in ordinary stars, the initial particles 1 and 2 are in a
dilute gas. In this case, the momentum-space densities may be replaced
by the Maxwell-Boltzmann forms,
\begin{equation}
n({\bf p}_1) = e^{-\beta ( E_1 - \mu_1)} \,, \qquad
n({\bf p}_2) = e^{-\beta ( E_2 - \mu_2)} \,.
\end{equation}
Expanding the Bose or Fermi denominators, we see that the correction to
this leading approximation involves the very small factors $ \exp\{
\beta \mu_1\} $ and $\exp \{ \beta \mu_2 \}$. Since in the dilute gas
limit
\begin{equation}
\langle n \rangle_\beta = \int { (d{\bf p} ) \over (2\pi)^3 } \, 
        e^{-\beta ( E - \mu ) } = \lambda^{-3} e^{ \beta \mu } \,,
\end{equation}
we see that the corrections to the dilute limit are of order
\begin{equation}
e^{\beta \mu} \approx \langle n \rangle_\beta \lambda^3 \,,
\label{mu}
\end{equation}
which, for example, at the core of the sun is roughly of order
$10^{-6}$. The corrections to the Maxwell-Boltzmann limit involve
exchange effects of quantum identical particles.  We turn now to
exploit the dilute nature of the initial reacting particles so that
they may be treated with classical statistics.

\subsection{Plasma}

We shall work out the nuclear reaction rate accounting exactly for
the Coulomb interactions between the reacting particles but taking
advantage of their assumed dilute character so that we may neglect
Bose or Fermi exchange effects. The other particles
in the plasma, however, --- such as electrons --- need not be dilute,
and they will not be restricted to obey classical statistics. 

We shall assume that the produced particles, which are labeled by 3
and 4, are sparsely distributed in the plasma so that their initial
state which appears in the rate formula (\ref{genresult}) thermal
expectation value $\langle \cdots \rangle_\beta$ can be represented by
the vacuum state. Thus the evaluation of the rate
involves the calculation of the vacuum expectation value in the
subspace of the produced particles
\begin{equation}
{\cal W}({\bf r}_3,{\bf r}_4,t) = \langle 0_3,0_4 | 
\psi_3({\bf r}_3,t) \, \psi_4({\bf r}_4,t) \, \psi^\dagger_4(0) \,
\psi^\dagger_3(0) | 0_3,0_4 \rangle 
\label{matrixel}
\end{equation}
in the limit ${\bf r}_3 = {\bf r} = {\bf r}_4$. The Heisenberg
equation of motion for the field operators gives
\begin{eqnarray}
i { \partial \over \partial t} \langle 0_3,0_4 | 
\psi_3({\bf r}_3,t) \, \psi_4({\bf r}_4,t) &=&
\left\{ - {\nabla^2_3 \over 2M_3} - {\nabla^2_4 \over 2M_4} + { e_3
e_4 \over |{\bf r}_4 - {\bf r}_3| } \right\} \langle 0_3,0_4 | 
\psi_3({\bf r}_3,t) \, \psi_4({\bf r}_4,t) 
\nonumber\\
&& \quad + \langle 0_3,0_4 | 
\psi_3({\bf r}_3,t) \, \psi_4({\bf r}_4,t) \Big[
e_3 \phi({\bf r}_3,t) 
+ e_4 \phi({\bf r}_4,t) \Big] \,.
\label{eqmot}
\end{eqnarray}
Here the operator $\phi({\bf r},t)$ is the potential produced by all the
other particles in the plasma. It is determined by the 
operator $\rho({\bf r}',t)$ which measures the charge density produced by
all these particles, 
\begin{equation}
\phi({\bf r},t) = \int (d{\bf r}') 
{ 1 \over | {\bf r} - {\bf r}' | } \rho( {\bf r}',t) \,.
\end{equation}
In the absence of this potential, the bra vector in Eq.~(\ref{eqmot}) 
defined by the field operators acting to the left on $\langle
0_3,0_4|$ obeys
\begin{equation}
i {\partial \over \partial t} \langle {\bf r}_3,{\bf r}_4 ,t| =
\langle {\bf r}_3,{\bf r}_4 ,t| H_{3,4} \,,
\end{equation}
where, with ordinary quantum mechanical notation,
\begin{equation}
H_{3,4} = { {\bf p}_3^2 \over 2M_3} + { {\bf p}_4^2 \over 2M_4}
        + { e_3 e_4 \over | {\bf r}_3 - {\bf r}_4 | } \,.
\end{equation}
Thus,
\begin{equation}
 \langle {\bf r}_3'= {\bf r} = {\bf r}_4' \,,\, t| =
\langle {\bf r}_3'= {\bf r} = {\bf r}_4' |
e^{-i H_{3,4} t} \,,
\end{equation}
The addition of the potential terms can be included by passing to an
interaction picture with the time evolution operator
\begin{equation}
W(t) = \left( \exp\left\{- i \int_0^t dt' \, [ 
e_3 \phi({\bf r}_3(t'),t') 
+ e_4 \phi({\bf r}_4(t'),t') ] \right\} \right)_+ \,,
\label{realt}
\end{equation}
a time-ordered exponential in the real time $t$ with
\begin{equation}
{\bf r}_{3 \, {\rm or} \, 4} \,(t) = 
e^{ iH_{3,4} t} \, {\bf r}_{3 \, {\rm or} \, 4} \,
                e^{-i H_{3,4} t } \,.
\end{equation} 
Thus
\begin{equation}
\langle 0_3,0_4 | \psi_3({\bf r}_3,t) \, \psi_4({\bf r}_4,t) =
\langle {\bf r}_3,{\bf r}_4 ,t| W(t) \,,
\label{braa}
\end{equation}
and we conclude that 
\begin{equation}
{\cal W}({\bf r}_3,{\bf r}_4,t) = \langle {\bf r}_3,{\bf r}_4 ,t | 
W(t) | {\bf r}_3 = 0 = {\bf r}_4 \rangle \,,
\end{equation}
since this obeys the equation of motion (\ref{eqmot}) and the correct
boundary conditions when $t \to 0$. The rate formula
(\ref{genresult}) is now expressed as
\begin{equation}
\Gamma =  g^2 \int_{-\infty}^{+\infty} dt \int (d{\bf r}) \, e^{iQt} 
\left\langle \psi^\dagger_1({\bf r},t) \, \psi^\dagger_2({\bf r},t) \,
{\cal W}({\bf r},{\bf r},t) \, \psi_2(0) \, \psi_1(0) 
\right\rangle_\beta \,.
\label{inter}
\end{equation}
It should be noted that ${\cal W}({\bf r},{\bf r},t)$ is an operator
in the plasma state space.

The thermal expectation value which appears in Eq.~(\ref{inter}) is
defined by a sum over all states (except for the final particle states
which have been accounted for), a trace weighted by the density
operator
\begin{equation}
{\cal P} = { 1 \over Z}  \exp\left\{ - \beta 
\left[ H - {\sum}_a \mu_a N_a \right] \right\}  \,,
\end{equation}
where $N_a$ are the number operators of the particles with chemical
potential $\mu_a$, and $Z$ is the partition function
\begin{equation}
Z = {\rm Tr} \exp\left\{ - \beta \left[H - {\sum}_a \mu_a N_a 
\right] \right\} \,.
\end{equation}
The density operator ${\cal P}$ defines the grand canonical ensemble
that we employ.  Since an exact treatment of the Coulomb interactions
of the particular reacting particles requires an emphasis on the
individual particle aspects, we are forced to employ the coordinate
representation for these particles. Thus we use the coordinate
representation for these particles in the trace that defines the
average in the grand canonical ensemble and write, in a necessarily
rather schematic way, 
\begin{eqnarray}
&&\left\langle \psi^\dagger_1({\bf r},t) \, \psi^\dagger_2({\bf r},t) \,
{\cal W}({\bf r},{\bf r},t) \, \psi_2(0) \, \psi_1(0) 
\right\rangle_\beta = \sum {1 \over N_1 !} 
{1 \over N_2 !} \int \cdots
\nonumber\\
&& \qquad\qquad 
\langle {\bf r}_1' \cdots \,;\, {\bf r}_2' \cdots \,;\,
 \cdots | {\cal P} \, 
\psi^\dagger_1({\bf r},t) \, \psi^\dagger_2({\bf r},t) \,
{\cal W}({\bf r},{\bf r},t) \, \psi_2(0) \, \psi_1(0) 
| {\bf r}_1' \cdots \,;\, {\bf r}_2' \cdots \,;\, \cdots \rangle \,.
\nonumber\\
&& \null
\end{eqnarray}
The sum is a sum over all particle numbers $N_1 \,, N_2 $
(which are now just numbers, not operators). 
The coordinates of all the particles that appear in the
state vectors are to be integrated over all space. The abbreviated
notation used to label these particles in a state vector such as
$$
| {\bf r}_1' \cdots \,;\, {\bf r}_2' \cdots \,; \cdots\rangle 
$$
is as follows. The first set $ {\bf r}_1' \cdots $ denotes the
coordinates of the $N_1$ particles of type 1 ; the second set 
$ {\bf r}_2' \cdots $ denotes the coordinates of the $N_2$ particles 
of type 2 ; the remaining $\cdots$ stand for all the other
particles. The sum defining the trace for these other particles may be
done in any representation. In view of the symmetry in the integration
over the initial reacting particles coordinates, the action of the
destruction and creation operators is, effectively
within this integration,
\begin{equation}
\psi_2(0) \, \psi_1(0) 
| {\bf r}_1' \cdots \,;\, {\bf r}_2' \cdots \,;\, \cdots \rangle 
 = \, | \cdots \,;\,  \cdots \,;\, \cdots \,;\, \rangle 
        N_1 N_2 \delta ({\bf r}_1') \delta ({\bf r}_2') \,.
\label{firstt}
\end{equation}
We now make the restriction that the initial reacting particles 1,2
are sufficiently dilute so that Boltzmann or classical statistics can
be employed. In this approximation, exchange effects are neglected,
and the particles that were destroyed
by the field operators in Eq.~(\ref{firstt}) are identified with the
particles destroyed when the remaining operators $\psi^\dagger_1({\bf
r},t)$ and $\psi^\dagger_2({\bf r},t)$ act to the left. 
Accordingly, the ensemble average which appears in the rate formula
(\ref{inter}) may be expressed as
\begin{eqnarray}
&&\left\langle \psi^\dagger_1({\bf r},t) \, \psi^\dagger_2({\bf r},t) \,
{\cal W}({\bf r},{\bf r},t) \, \psi_2(0) \, \psi_1(0) 
\right\rangle_\beta =
 \sum {1 \over (N_1-1) !} {1 \over (N_2-1) !} 
\int \cdots
\nonumber\\
&& \qquad\qquad
\langle  \cdots \,;\, \cdots \,;\, \cdots | \,
\langle {\bf r}_1' = 0 = {\bf r}_2' | {\cal P} 
\psi^\dagger_1({\bf r},t) \, \psi^\dagger_2({\bf r},t) \,
{\cal W}({\bf r},{\bf r},t) \, | 0_1,0_2 \rangle 
\, |\cdots \,;\, \cdots \,;\, \cdots \rangle 
\nonumber\\
&& \qquad\qquad\qquad = {\rm Tr} 
 \langle {\bf r}_1' = 0 = {\bf r}_2' | {\cal P}
\psi^\dagger_1({\bf r},t) \, \psi^\dagger_2({\bf r},t) \,
 | 0_1,0_2 \rangle {\cal W}({\bf r},{\bf r},t) \,.
\label{advent}
\end{eqnarray}

The action of the creation operators on the vacuum state
$|0_1,0_2\rangle$ which appears here is essentially the adjoint of
that appearing in Eq.~(\ref{braa}), and so we have
\begin{equation}
\psi^\dagger_1({\bf r},t) \, \psi^\dagger_2({\bf r},t) \,
 | 0_1,0_2 \rangle = V(t) | {\bf r}_1' = {\bf r} = {\bf r}_2' \,, t
\rangle \,.
\end{equation}
Here the time dependence of the state is governed by the
quantum-mechanical Hamiltonian
\begin{equation}
H_{1,2} = { {\bf p}_1^2 \over 2M_1} + { {\bf p}_2^2 \over 2M_2}
        + { e_1 e_2 \over | {\bf r}_1 - {\bf r}_2 | } \,,
\end{equation}
with
\begin{equation}
| {\bf r}_1'= {\bf r} = {\bf r}_2' \,,\, t \rangle = 
e^{ i H_{1,2} t} | {\bf r}_1'= {\bf r} = {\bf r}_2'  \rangle \,,
\end{equation}
and
\begin{equation}
V (t) = \left( \exp\left\{ i \int_0^t dt' \, [ 
e_1 \phi({\bf r}_1(t'),t') 
+ e_2 \phi({\bf r}_2(t'),t') ] \right\} \right)_- 
\end{equation}
is an anti-time-ordered exponential in the real time $t$ with
\begin{equation}
{\bf r}_{1 \, {\rm or} \, 2}\,(t) = 
e^{ iH_{1,2} t} \, {\bf r}_{1\, {\rm or} \,2} \,
                e^{-i H_{1,2} t } \,.
\end{equation}
To deal with the action of the statistical density operator ${\cal
P}$, we write the total Hamiltonian for the whole system which appears
in ${\cal P}$ as
\begin{equation}
H = H_{1,2} + e_1 \phi({\bf r}_1) + e_2 \phi({\bf r}_2) + H_{\rm R}
\,,
\end{equation}
where $\phi({\bf r})$ is the potential that the remaining particles in
the plasma produce at the point ${\bf r}$, and $H_{\rm R}$ is the
Hamiltonian for the remaining particles in the plasma. In this way, we
can make use of the interaction picture in imaginary time in writing
\begin{equation}
\exp\{ -\beta H \} = \exp\{ -\beta (H_{1,2} + H_{\rm R}) \} \, U(\beta)
\,,
\label{write}
\end{equation}
in which
\begin{equation}
U(\beta) = \left( \exp\left\{ - \int_0^\beta d\tau [ 
e_1 \phi({\bf r}_1(\tau),\tau) 
+ e_2 \phi({\bf r}_2(\tau),\tau) ] \right\} \right)_+ \,,
\label{intpic}
\end{equation}
is a time-ordered exponential in the imaginary time $\tau$ with
\begin{equation}
{\bf r}_{1\, {\rm or}\, 2}\, (\tau) = e^{H_{1,2} \tau} \, 
{\bf r}_{1 \, {\rm or} \,  2} \, e^{-H_{1,2} \tau } \,.
\end{equation}
The explicit $\tau$ dependence in the potential terms appears because
the charge density of the background plasma moves in imaginary time
under the action of $H_{\rm R}$. That is,
\begin{equation}
\phi({\bf r}, \tau) = \int (d{\bf r}') 
{ 1 \over | {\bf r} - {\bf r}' | } \rho( {\bf r}' , \tau) \,,
\end{equation}
where
\begin{equation}
\rho( {\bf r}' , \tau) = e^{ H_{\rm R} \tau } \rho( {\bf r}' )
                e^{ - H_{\rm R} \tau } \,.
\end{equation}
Using Eq.~(\ref{write}), we explicitly separate out the initial
particles from the density operator to obtain 
\begin{equation}
{\cal P} = e^{\beta(\mu_1 + \mu_2)} \bar{\cal P} e^{-\beta H_{1,2} }
U(\beta) \,,
\end{equation}
where $\bar{\cal P}$ is the density operator for the remaining
particles. Hence with
\begin{equation}
\langle {\bf r}_1'=0 = {\bf r}_2' | e^{ -\beta H_{1,2} } 
= \langle {\bf r}_1'=0 = {\bf r}_2' \,,\ -i\beta | \,,
\end{equation} 
we have 
\begin{eqnarray}
&& \langle {\bf r}_1' = 0 = {\bf r}_2' | {\cal P}
\psi^\dagger_1({\bf r},t) \, \psi^\dagger_2({\bf r},t) \,
 | 0_1,0_2 \rangle = 
\nonumber\\
&& \qquad\qquad
\exp\{\beta(\mu_1 + \mu_2)\} \, \bar{\cal P} \, 
 \langle {\bf r}_1' = 0 = {\bf r}_2' \,, -i\beta | 
U(\beta) V(t) | {\bf r}_1' = {\bf r} = {\bf r}_2' \,, t \rangle \,.
\end{eqnarray}
Inserting this result in Eq.~(\ref{advent}) produces a plasma thermal
expectation value, and so Eq.~(\ref{inter}) yields
\begin{eqnarray}
\Gamma = &g^2& \exp\{ \beta ( \mu_1 + \mu_2 )\}
\int_{-\infty}^{+\infty} dt \int (d{\bf r}) \, e^{iQt}
\nonumber\\
&& \qquad  \Big\langle
        \langle {\bf r}_1'=0 = {\bf r}_2' \,,\, -i\beta | 
        U(\beta) V (t)
        | {\bf r}_1'= {\bf r} = {\bf r}_2' \,,\, t \rangle
\nonumber\\
&&  \qquad \qquad
\langle {\bf r}_3'= {\bf r} = {\bf r}_4' \,,\, t| W(t)
|{\bf r}_3'=0 ={\bf r}_4'  \rangle  \Big\rangle_\beta \,. 
\label{great}
\end{eqnarray}
This expression for the reaction rate is the major result of this
paper. It separates out the plasma interactions with the particles
undergoing the nuclear reaction from their internal Coulomb
interactions. The plasma interactions have two different effects. One
is to alter the thermal distribution describing the density
of states of the initial two particles that enter into the
reaction. This alteration is described by the operator $U(\beta)$. The
other type of effect of the background plasma is to alter the
dynamics of the motion of the particles undergoing the nuclear
reaction. This is described by the operators $V(t)$ and $W(t)$.

Since our notation is rather condensed, it is worthwhile describing
the meaning of this result in a little more detail. Spelling out the
time dependence of the states expresses
\begin{eqnarray}
\Gamma = & g^2 & \exp\{ \beta ( \mu_1 + \mu_2) \}
\int_{-\infty}^{+\infty} dt \int (d{\bf r}) e^{iQt}
\nonumber\\
&& \qquad  \Big\langle
        \langle {\bf r}_1'=0 = {\bf r}_2' | 
        e^{ -\beta H_{1,2} } U(\beta) V (t) e^{ i H_{1,2} t}
        | {\bf r}_1'= {\bf r} = {\bf r}_2'  \rangle
\nonumber\\
&& \qquad \qquad
\langle {\bf r}_3'= {\bf r} = {\bf r}_4' |
e^{-i H_{3,4} t} W(t)
|{\bf r}_3'=0 ={\bf r}_4'  \rangle  \Big\rangle_\beta \,. 
\end{eqnarray}
Consider, for
example, the second matrix element that appears within the thermal
expectation value. Writing out the explicit 
expression (\ref{realt}) for the operator $W(t))$, this is the matrix 
element
\begin{eqnarray}
&& \langle {\bf r}_3'= {\bf r} = {\bf r}_4' \,,\,  t| W(t)
|{\bf r}_3'=0 ={\bf r}_4'  \rangle  =
\nonumber\\
&& \,\, \langle {\bf r}_3'= {\bf r} = {\bf r}_4' \,,\, t|
\left( \exp\left\{- i \int_0^t dt' \, [ 
e_3 \phi({\bf r}_3(t'),t') 
+ e_4 \phi({\bf r}_4(t'),t') ] \right\} \right)_+
|{\bf r}_3'=0 ={\bf r}_4'  \rangle \,.
\label{example}
\end{eqnarray}
The potential $\phi({\bf r}_3(t'),t')$ is a field operator in the
space of states that describe the thermal ensemble denoted by the
overall expectation value $\langle \cdots \rangle_\beta$. This field
operator is evaluated at the position ${\bf r}_3(t')$ which itself is
an operator in the quantum-mechanical two-particle states of the final
produced particles. The expectation value of Eq.~(\ref{example}),
which is in the two-particle space of the produced particles, yields a
functional of the field operators $\phi$ that depends parametrically
upon the coordinate ${\bf r}$ and the time $t$. A similar functional
of the field operator $\phi$ is produced by the other, initial
particle, matrix element in Eq.~(\ref{great}). The reaction rate
$\Gamma$ is then obtained by the space-time integral of the plasma,
thermal expectation value of these functionals as shown in
Eq.~(\ref{great}).

In solar fusion processes, the energy release $Q$ of the nuclear
reaction is much larger than the temperature $T$. In such cases, the
integrand in the time integration in Eq.~(\ref{great}) is dominated by
the rapidly varying phase factor $\exp\{ iQt \}$ and the time
dependence of the states 
$\langle {\bf r}_3' = {\bf r} = {\bf r}_4' \,,\, t |$. 
Including only this time variation, and setting $t=0$ elsewhere in
Eq.~(\ref{great}), the time integration produces an energy-conserving
$\delta$ function that sets the energy of the produced particles equal
to the energy release $Q$.  The time region that is important is of
the order $1/Q$, and the leading terms in the large $Q$ limit are
obtained by placing $t=0$ in $V(t)$ and $W(t)$, which are thus
replaced by unity, and by placing $t=0$ in the state $ | {\bf r}_1' =
{\bf r} = {\bf r}_2' \,,\, t \rangle$ as well, with
corrections\footnote{In the cases that we are considering, the thermal
excitations in the plasma provide energy transfers of order $T$ which
is much less than $Q$. However, in the interiors of stars that are
more highly evolved than the sun, less exothermic and even endothermic
reactions become important, and then the full complexity of the time
dependence in Eq.~(\ref{great}) must be taken into account.}
to this approximation roughly of order $ T / Q$. We shall see this
explicitly in later examples. Thus in the highly exothermic limit, the
plasma dependence in the general rate formula (\ref{great}) enters
only in the quantity
\begin{eqnarray}
&& \Big\langle \langle {\bf r}_1'= {\bf r} = {\bf r}_2' \,,\, -i\beta | 
  U(\beta) | {\bf r}_1'= 0 = {\bf r}_2'  \rangle \Big\rangle_\beta =
\nonumber\\ 
&& \qquad \langle {\bf r}_1'=0 = {\bf r}_2' | \exp\{- \beta H_{1,2} \} 
\nonumber\\
&& \quad
\left\langle \left( \exp\left\{ - \int_0^\beta d\tau [ 
e_1 \phi({\bf r}_1(\tau),\tau) 
+ e_2 \phi({\bf r}_2(\tau),\tau) ] \right\} \right)_+
\right\rangle_\beta
        | {\bf r}_1'= {\bf r} = {\bf r}_2' \rangle \,.
\label{probb}
\end{eqnarray}
The spatial coordinate integration over ${\bf r}$ in
Eq.~(\ref{great}) produces momentum $\delta$ functions that enforce
the conservation of the total momentum in the nuclear reaction. As we
shall also describe later, the effect of this momentum conservation is
to replace the ion masses in the plasma that have Coulomb
interactions with the fusing nuclei by proper reduced masses that
include the total mass of these nuclei. This takes into account 
the center-of-mass motion of the initial nuclear system through the
plasma. 

\section{Rate Calculations}

\subsection{Dilute Limit}

We first turn to illustrate the character of our basic result
(\ref{great}) with the limit where the plasma is very dilute. In this
limit, the Coulomb interactions of the reacting particles with the
background plasma may be neglected, and thus the operators $U(\beta)$,
$V (t)$, and $W(t)$ may be replaced by the identity operator. Thus in
this dilute limit, the thermal expectation value is the trivial
expectation value of the identity operator, and our result becomes 
simply
\begin{eqnarray}
\Gamma_C = &g^2& \exp\{ \beta ( \mu_1 + \mu_2 )\}
\int_{-\infty}^{+\infty} dt \int (d{\bf r}) e^{iQt}
\nonumber\\
&& \qquad 
        \langle {\bf r}_1'=0 = {\bf r}_2' , -i\beta | 
        {\bf r}_1'= {\bf r} = {\bf r}_2' , t \rangle
\langle {\bf r}_3'= {\bf r} = {\bf r}_4' , t|
        {\bf r}_3'=0 ={\bf r}_4'  \rangle  \,. 
\label{coulb}
\end{eqnarray}
To evaluate the transformation functions that appear here, we
introduce complete sets of intermediate states which are eigenstates 
of the total initial or final momentum, ${\bf P}$ or ${\bf P}'$, and 
of the initial or final relative momentum, ${\bf p}$ or ${\bf
p}'$. The relative momenta are the asymptotic momenta of the
interacting scattering states. Thus
\begin{eqnarray}
\langle {\bf r}_1'=0 = {\bf r}_2' , -i\beta | 
        {\bf r}_1'= {\bf r} = {\bf r}_2' , t \rangle &=&
\int { (d {\bf P}) \over (2\pi)^3} 
\exp\left\{ - { {\bf P}^2 \over 2M } \, (\beta - it) - i {\bf P} \cdot
{\bf r} \right\} 
\nonumber\\
&& 
\int { (d {\bf p}) \over (2\pi)^3} 
\exp\left\{ - { {\bf p}^2 \over 2m }\, (\beta - it) \right\}
\left| \psi^{(i)}_{\bf p}(0) \right|^2 \,,
\end{eqnarray}
and
\begin{eqnarray}
\langle {\bf r}_3'= {\bf r} = {\bf r}_4' , t|
        {\bf r}_3'=0 ={\bf r}_4'  \rangle  &=&
\int { (d {\bf P}') \over (2\pi)^3} 
\exp\left\{ -i { {{\bf P}'}^2 \over 2M } \, t + i {\bf P}' \cdot
{\bf r} \right\} 
\nonumber\\
&& \quad 
\int { (d {\bf p}') \over (2\pi)^3} 
\exp\left\{ -i { {{\bf p}'}^2 \over 2m' } \, t \right\}
\left| \psi^{(f)}_{{\bf p}'}(0) \right|^2 \,.
\end{eqnarray}
Here
\begin{equation}
M = M_1 + M_2 = M_3 + M_4 
\end{equation}
is the common total mass of the initial and final states, and
\begin{equation}
{1 \over m} = {1 \over M_1} + {1 \over M_2} \,,\, \qquad
{1 \over m'} = {1 \over M_3} + {1 \over M_4}
\end{equation}
are the reduced masses of the initial and final states in the
reaction. The squared amplitudes $|\psi_{{\bf p}}(0)|^2 $ are the
squared relative-motion Coulomb wave functions at the origin which
have the form
\begin{equation}
|\psi_{{\bf p}}(0)|^2 = { 2\pi \eta e^{-2\pi \eta} \over 1 - e^{- 2\pi
\eta} } \,,
\label{gammow}
\end{equation}
with
\begin{equation}
\eta = { e_1 e_2 \, m \over |{\bf p}| } \,,\ \qquad
\eta '= { e_3 e_4 \, m' \over |{\bf p}'| } \,,
\end{equation}
for the initial and final states. 

The space-time integration in Eq.~(\ref{coulb}) now produces
energy-momentum conserving $\delta$ functions, and we have
\begin{eqnarray}
\Gamma_C = &g^2& \exp\{ \beta ( \mu_1 + \mu_2) \}
\int { (d {\bf P}) \over (2\pi)^3} 
\exp\left\{ - \beta \, { {\bf P}^2 \over 2M } \right\} 
\int { (d {\bf p}) \over (2\pi)^3} 
\exp\left\{ - \beta \, { {\bf p}^2 \over 2m } \right\} 
\left| \psi^{(i)}_{\bf p}(0) \right|^2
\nonumber\\
&&  \int { (d {\bf P}') \over (2\pi)^3} 
\int { (d {\bf p}') \over (2\pi)^3} 
\left| \psi^{(f)}_{{\bf p}'}(0) \right|^2
(2\pi)^3 \delta ( {\bf P}' - {\bf P} ) 
(2\pi) \delta \left( { {{\bf p}'}^2 \over 2m' } -
        { {\bf p}^2 \over 2m }  - Q \right) \,.
\end{eqnarray}
Since $Mm = M_1 M_2$,
\begin{equation}
\lambda^{-3}_1 \lambda^{-3}_2 = 
\left( { M_1 \over 2\pi \beta } \right)^{3/2} 
\left( { M_2 \over 2\pi \beta } \right)^{3/2}
= \left( { M \over 2\pi \beta } \right)^{3/2} 
\left( { m \over 2\pi \beta } \right)^{3/2} \,,
\end{equation}
and remembering the formula (\ref{mu}) for the number density, we see
that
\begin{eqnarray}
\Gamma_C &=& \langle n_1 \rangle_\beta \langle n_2 \rangle_\beta 
\, g^2 \left( { \beta \over 2\pi m } \right)^{3/2} 
\int (d {\bf p})
\exp\left\{ - \beta \, { {\bf p}^2 \over 2m } \right\} 
\left| \psi^{(i)}_{\bf p}(0) \right|^2
\nonumber\\
&& \qquad 
\int { (d {\bf p}') \over (2\pi)^3} 
\left| \psi^{(f)}_{{\bf p}'}(0) \right|^2
(2\pi) \delta \left( { {{\bf p}'}^2 \over 2m' } -
        { {\bf p}^2 \over 2m }  - Q \right) \,.
\label{coulrate}
\end{eqnarray}
This is of the form
\begin{equation}
\Gamma_C = \langle n_1 \rangle_\beta \langle n_2 \rangle_\beta
        \int \langle d\sigma v \rangle_\beta \,,
\end{equation}
where now the differential cross section $d\sigma$ for the reaction
includes the Coulomb corrections for the initial and final states.

Before turning to the plasma corrections, let us examine the scale ---
the size --- of the variables that enter into the reaction rate
calculation. Clearly, the common center-of-mass momentum for the
initial and final states is governed by the Boltzmann factor. Thus, in
order-of-magnitude,
\begin{equation}
P = P' \sim \sqrt{ { 2M \over \beta } } \,.
\label{Pest}
\end{equation}
The corresponding extent of the motion of the
center of mass in imaginary time is of order
\begin{equation}
\Delta R \sim (P/M) \beta \sim \sqrt{ \beta \over M } \sim \Lambda \,,
\end{equation}
where $\Lambda$ is the thermal wavelength of the center-of-mass
motion.  
The magnitude of the initial relative momentum is determined by the
balance of two factors, the Boltzmann factor which rapidly decreases
as this momentum increases and the Coulomb barrier factor whose effect
weakens as the relative momentum increases. In the usual applications,
$2\pi \eta = 2\pi e_1 e_2 \, m / p $ is a moderately large number 
in the momentum region of relevance, and the value (\ref{gammow}) of 
the Coulomb wave function at the origin is dominated by the leading 
exponential so that the distribution of initial relative momentum is 
peaked at the minimum value of the exponent in the factor
\begin{equation}
\exp\left\{ - \beta { p^2 \over 2m } - 2\pi { e_1 e_2 \, m \over p
} \right\} \,,
\label{exp}
\end{equation}
which is the point $ p = \bar p$ given by
\begin{equation}
\bar p^3 = 2\pi e_1 e_2 \, m^2 / \beta \,.
\label{probp}
\end{equation}
At this point, the Coulomb parameter has the value
\begin{equation}
\bar \eta = { e_1 e_2 \, m \over \bar p} = \left( { e_1^2 e_2^2 
\, m \, \beta \over 2\pi } \right)^{1/3} \,,
\label{coul}
\end{equation}
and the exponential displayed in Eq.~(\ref{exp}) becomes $ \exp\{- 3\pi
\bar \eta \}$. Note that
\begin{equation}
\lambda_{12} = \sqrt { 2\pi \beta \over m}
\end{equation}
defines a thermal wavelength, while
\begin{equation}
a_{12} = {1 \over e_1 e_2 \, m}
\end{equation}
defines the Bohr radius of the initial reacting system. In terms of
these parameters,
\begin{equation}
\bar p^3 = { (2\pi)^2 \over \lambda_{12}^2 \, a_{12} } \,.
\label{pbarr}
\end{equation}
The corresponding kinetic energy $ \bar p^2 / 2m $ is, in general,
much less than the energy release $Q$ of the nuclear reaction. In
these cases, the relative momentum of the produced particles is, in
view of the energy-conserving $\delta$ function in the rate
(\ref{coulrate}), of order
\begin{equation}
p' \sim \sqrt{ 2 m' Q} \,, 
\label{pest}
\end{equation}
which is much larger than the typical initial relative momentum $ \bar p$.

The value (\ref{gammow}) of the Coulomb wave function at the origin is
given, to leading order, by a simple tunneling process. The classical
turning point $r = r_{\rm max}$ which enters here, at the typical
initial relative momentum $\bar p$, is determined by
\begin{equation}
{ \bar p^2 \over 2m } - {e_1 e_2 \over r_{\rm max} } = 0 \,,
\end{equation}
or
\begin{equation}
r_{\rm max} = { 2 e_1 e_2 \, m \over \bar p^2} = { 2 \bar \eta \over
\bar p } \,.
\label{turn}
\end{equation}
This sets the scale of the quantum motion in imaginary time for
$r(\tau)$. Indeed, the tunneling amplitude for the initial
two-particle state in a thermal ensemble at temperature $ T = 1/\beta$
can be obtained as the steepest descent approximation to the path
integral representation [Alastuey and Jancovici (1978)]. In this
approximation, the exponential $\exp\{ - 3\pi \bar \eta \}$ appears as
the exponent of the classical action computed for a classical solution
in imaginary time that has an orbit extending from $r=0$ to $r = r_{\rm
max}$ back to $r=0$. This classical solution gives the leading
approximation to the quantum motion.

\subsection{Plasma Corrections}

As we remarked in the discussion of the general rate formula
(\ref{great}), for highly exothermic reactions we may neglect the
effects of the operators $V(t)$ and $W(t)$ so that the only plasma
dependence is contained in the operator $U(\beta)$. The scale of
spatial variations in the plasma is controlled by the Debye length
$\kappa_D^{-1}$ which is much larger than the estimates of the size of
the coordinates ${\bf r}_{1 \, {\rm or} \, 2}(\tau) $ which was just
presented. Hence, we may, in the first approximation, set ${\bf r} =
0$ in the potential operator $\phi({\bf r}(\tau),\tau)$. As we shall
soon see, the imaginary time dependence in $\phi({\bf r},\tau)$ may
also be neglected in leading order so that we may approximate 
\begin{equation}
U(\beta) = \exp\{ -\beta (e_1+e_2) \phi(0) \} \,.
\end{equation}
Assuming that the plasma ensemble may be described by uncorrelated,
Gaussian statistics, we now have
\begin{equation}
\big\langle U(\beta) \big\rangle_\beta = 
\exp\{  \beta^2 (e_1+e_2)^2 (1/2) \langle \phi(0) \phi(0)\rangle_\beta
\} \,.
\end{equation}
Using the potential correlation function obtained in Appendix A,
Eq.~(\ref{phiclass}), 
\begin{equation}
\big\langle \phi({\bf r}) \phi({\bf r}') \big\rangle_\beta
= {1 \over \beta |{\bf r} - {\bf r}'| } \left[1 - \exp\{ - \kappa_D
        |{\bf r} - {\bf r}'| \}  \right] \,,
\label{pot}
\end{equation}
we have
\begin{equation}
\big\langle U(\beta) \big\rangle_\beta = 
\exp\{ (1/2) \beta (e_1+e_2)^2 \kappa_D \} \,.
\end{equation}
It should be remarked that the exponent that appears here is
generally not a large number so that the Gaussian statistics
assumption is not needed. As remarked in the Introduction, factors
combine to form the plasma corrected number densities, and as shown in
Appendix B (assuming that the result exponentiates)
\begin{equation}
\big\langle n_{1 \, {\rm or} \, 2} \big\rangle_\beta
= \lambda_{1 \, {\rm or} \, 2}^{-3} \exp\{ \beta \mu_{1 {\rm or} \, 2} \}
\exp\{ (1/2) \beta e_{1 \, {\rm or} \, 2}^2 \kappa_D \} \,.
\end{equation}
Therefore, the leading plasma correction is given by
\begin{equation}
\Gamma_S = \exp\{ \beta e_1 e_2 \, \kappa_D \} \Gamma_C \,,
\end{equation}
where $\Gamma_C$ is the Coulomb corrected rate (\ref{coulrate}). And,
as remarked in the Introduction, this is the Salpeter corrected rate.

We turn now to examine the size of the terms that we have neglected.
We consider first the ``dynamical'' correction. Since the important
time region is the small time interval $0 < |t| < 1/Q$, in which  
the reacting particles move little, in the first approximation we may
neglect both the spatial and time coordinates in the potential
operator $\phi({\bf r},t)$.  Hence,
in leading order, the dynamical corrections involve
\begin{equation}
V(t) W(t) = \exp\{ it (e_1 + e_2) \phi(0) \} 
\exp\{ -it (e_3 + e_4) \phi(0) \} = 1 \,,
\end{equation}
since charge is conserved in the reaction, $e_1 + e_2 = e_3 + e_4$. 
To obtain the first ``dynamical correction'', we note that since short
times are involved, the dominant term is that with the least number of
time integrations. This term involves the interference of a correction
from the ``statistical'' $U(\beta)$ and one of the ``dynamical''
$V(t)$ or $W(t)$ factors,
\begin{equation}
\pm i \beta (e_1+e_2) e_r \int_0^t dt' \Big\langle
\phi(0) [\phi({\bf r}(t'),t') - \phi(0)]
\Big\rangle_\beta \,.
\label{correctt}
\end{equation}
Here the $\pm$ is $+$ for the $V(t)$ factor of the initial particles
$1,2$ and $-$ for the $W(t)$ factor of the final particles $3,4$. 
The charge of one of the reacting particles is denoted by $e_r$ and ${\bf
  r}(t')$ is its position operator. This correction is examined in
Appendix E which provides the estimate
\begin{equation}
(\beta e_1e_2 \, \kappa_D) \left\{ 
{ \hbar \omega_p \over Q} \sqrt{ { 1 \over \beta Q} } \right\} \,, 
\end{equation}
where $\omega^2_p \sim \kappa^2_D / \beta M $ defines the plasma
frequency of a typical ion in the plasma. We have reverted to ordinary
units in displaying this factor with Planck's constant $\hbar$ to show
explicitly that it is of a quantum mechanical nature.  The factor in
braces is generally a very small correction indeed to the basic
Salpeter correction $(\beta e_1e_2 \, \kappa_D)$. For example, at the
center of the sun, and for $Q =1$ MeV, the factor in braces is about
$10^{-6}$. Although one might expect that the contribution of the
electrons in the plasma with their larger plasma frequency would
dominate the dynamical correction, as shown in Appendix E, the leading
contribution of the electrons cancels when summed over all the
reacting particles. 

We turn at last to estimate the error involved in the leading rate
correction which comes from the ``statistical'' $U(\beta)$. 
(A related and simpler calculation of the single-particle number
density is presented in Appendix B.)
To make this estimate, we examine the initial particle expectation value
of the dilute plasma limit
\begin{eqnarray}
&&  \Big\langle
        \langle {\bf r}_1'=0 = {\bf r}_2' \,,\, -i\beta | 
         U(\beta)
        | {\bf r}_1'= {\bf r} = {\bf r}_2' \,,\, t \rangle
        \Big\rangle_\beta
\nonumber\\
&& \quad \simeq 
        \langle {\bf r}_1'=0 = {\bf r}_2' \,,\, -i\beta | \Bigg\{
    1 + {1\over2} \int_0^\beta d\tau \, d\tau' \, 
\nonumber\\
&& \qquad 
\Big\langle \Big(
        [ e_1 \phi({\bf r}_1(\tau),\tau) 
        + e_2 \phi({\bf r}_2(\tau),\tau) ] 
        [ e_1 \phi({\bf r}_1(\tau'),\tau') 
        + e_2 \phi({\bf r}_2(\tau'),\tau') ] \Big)_+
        \Big\rangle_\beta  \Bigg\}
\nonumber\\
&& \qquad \qquad \qquad \qquad \qquad
        | {\bf r}_1'= {\bf r} = {\bf r}_2' \,,\, t \rangle \,.
\label{correct}
\end{eqnarray}
The plasma expectation value that appears here may be written as a
spatial Fourier transform
\begin{equation}
\Big\langle \Big( \phi({\bf r},\tau) 
        \phi({\bf r}',\tau') \Big)_+ \Big \rangle_\beta
        = \int { (d {\bf k}) \over (2\pi)^3 } e^{ i {\bf k} \cdot
                ( {\bf r} - {\bf r}' ) } G_\beta ( k ,\, \tau 
                - \tau' ) \,.
\label{fourier}
\end{equation}
To discuss the initial particle expectation value, we use the
center-of-mass and relative variables, 
\begin{equation}
{\bf R} = ( M_1 \, {\bf r}_1 + M_2 \, {\bf r}_2 ) / M \,, \qquad 
{\bf P} = {\bf p}_1 + {\bf p}_2 \,,
\end{equation}
and
\begin{equation}
\check{\bf r} = {\bf r}_1 - {\bf r}_2 \,, \qquad {\bf p} = 
        (M_2 \, {\bf p}_1 - M_1 \, {\bf p}_2 ) / M \,,
\end{equation}
where, as before, $M = M_1 + M_2$ is the total mass of the reacting
system. In terms of these variables this expectation value has the
direct product form
\begin{eqnarray}
 \Big\langle {\bf r}_1'= 0 = {\bf r}_2' \,,\, -i\beta && \Big| \cdots
   \Big| {\bf r}_1'= {\bf r} = {\bf r}_2' \,,\, t \Big\rangle 
\nonumber\\
 &=&  \Big\langle {\bf R}' = 0 \Big| \exp\left\{ -\beta { {\bf P}^2 \over
        2M} \right\} \Big\langle \check{\bf r}' = 0 \Big| 
                \exp\left\{ -\beta H_r \right\} \cdots
\nonumber\\
&& \qquad \qquad 
         \exp\left\{ i { {\bf P}^2 \over 2M} t \right\} \Big| 
                {\bf R}' = {\bf r} \Big\rangle
        \exp\left\{ i H_r t \right\} \Big| \check{\bf r}' = 0 
                \Big\rangle \,,
\end{eqnarray}
in which
\begin{equation}
H_r = { {\bf p}^2 \over 2m } + { e_1 e_2 \over |\check{\bf r}| } 
\end{equation}
is the Hamiltonian for the relative motion of the initial system with
reduced mass $m$.

We shall first work out the effects of the center-of-mass motion. 
In terms of the center-of-mass and relative coordinates,
\begin{equation}
{\bf r}_1 = {\bf R} + M_2 \, \check{\bf r} \, / \, M \,, \qquad
{\bf r}_2 = {\bf R} - M_1 \, \check{\bf r} \, / \, M \,.
\end{equation}
Thus, using the Fourier transform (\ref{fourier}) in evaluating the
correction (\ref{correct}), we see that we need to calculate the time
ordered product 
$$
\Big( \exp\{ i {\bf k} \cdot [ {\bf R}(\tau) - {\bf R}(\tau') ]
        \} \Big)_+ \,.
$$
The center-of-mass coordinate undergoes free particle motion in
imaginary time,
\begin{equation}
{\bf R}(\tau) = {\bf R} - i {\bf P} \tau / M \,.
\end{equation}
Using the canonical commutation relations to order the exponential,
\begin{eqnarray}
\exp\{ i {\bf k} \cdot {\bf R} + {\bf k} \cdot {\bf P} \tau / M \}
        &=& \exp\{ {\bf k}^2 \tau / 2M \} \exp\{ i {\bf k} \cdot
                {\bf R} \} \exp\{ {\bf k} \cdot {\bf P} \tau /M \}
\nonumber\\
        &=& \exp\{ - {\bf k}^2 \tau / 2M \} 
                \exp\{ {\bf k} \cdot {\bf P} \tau /M \}
                \exp\{ i {\bf k} \cdot {\bf R} \} \,,
\end{eqnarray}
we find that
\begin{equation}
\Big( \exp\{ i {\bf k} \cdot [ {\bf R}(\tau) - {\bf R}(\tau') ]
        \} \Big)_+ = \exp\{- {\bf k}^2 |\tau - \tau'| / 2M \} 
        \exp\{ {\bf k} \cdot {\bf P} (\tau - \tau') /M \} \,.
\end{equation}
With this result in hand, we can compute explicitly the center-of-mass
contribution to the initial particle expectation value needed for the
correction (\ref{correct}). Introducing a complete set of intermediate
state total momentum eigenstates, we now have
\begin{eqnarray}
 \Big\langle {\bf R}' = 0 \Big| \exp\left\{ -\beta { {\bf P}^2 \over
        2M} \right\}
\Big( &\exp&\{ i {\bf k} \cdot [ {\bf R}(\tau) - {\bf R}(\tau') ]
        \} \Big)_+
 \exp\left\{ i { {\bf P}^2 \over 2M} t \right\} \Big| 
                {\bf R}' = {\bf r} \Big\rangle
\nonumber\\
         &=& \exp\{- {\bf k}^2 |\tau - \tau'| / 2M \} 
                \int { (d {\bf P}) \over (2\pi)^3 }
                \exp\{ -i {\bf P} \cdot {\bf r} \}
\nonumber\\
        && \quad 
        \exp\left\{ - { {{\bf P}}^2 \over 2M} ( \beta -it) \right\} 
        \exp\{ {\bf k} \cdot {\bf P} (\tau - \tau') /M \} \,.
\label{center}
\end{eqnarray}
The spatial $(d{\bf r})$ integration which appears in the rate formula
(\ref{great}) produces a $\delta$ function which identifies the
initial and final center-of-mass momenta ${\bf P}$ and ${\bf
P}'$. Thus the time-dependent factor 
$ \exp\left\{  i { {{\bf P}}^2 \over 2M} \, t \right\} $
in Eq.~(\ref{center}) cancels the factor
$ \exp\left\{  - i { {\bf P}^2 \over 2M} \, t \right\} $
which is contained in the final produced particle matrix
element. Thus, effectively, we may delete all the time dependence
associated with the center-of-mass motion, delete the ${\bf r}$
dependence, and remove the integration over the center-of-mass
momentum ${\bf P}'$ of the final produced particles. With this
understood, we may complete the square in the $(d{\bf P})$ integration
in Eq.~(\ref{center}) and get, effectively,
\begin{eqnarray}
 \Big\langle {\bf R}' = 0 \Big| \exp\left\{ -\beta { {\bf P}^2 \over
        2M} \right\} 
\Big( &\exp&\{ i {\bf k} \cdot [ {\bf R}(\tau) - {\bf R}(\tau') ]
        \} \Big)_+
 \exp\left\{ i { {\bf P}^2 \over 2M} t \right\} \Big| 
                {\bf R}' = {\bf r} \Big\rangle
\nonumber\\
         &=& \left( {M \over 2\pi \beta} \right)^{3/2}
                C( k \,,\, \tau - \tau' ) \,,
\label{centerb}
\end{eqnarray}
in which 
\begin{equation}
C( k \,,\, \tau - \tau' ) =
\exp\left \{- {\beta {\bf k}^2 \over 2M } f(\tau - \tau')
\right\}  \,,
\label{centerc}
\end{equation}
where
\begin{equation}
f(\tau - \tau') = { |\tau - \tau'| \over \beta } \left( 1 - 
{ | \tau - \tau'| \over \beta } \right) \,.
\end{equation}

In an attempt to restrain notational proliferation, we define the
normalized expectation value of an operator $X$ by
\begin{equation}
\Big(\Big| \, X \, \Big|\Big) = {
  \Big\langle \check{\bf r}' = 0 \Big| 
                \exp\left\{ -\beta H_r \right\} \, X \,
        \exp\left\{ i H_r t \right\} \Big| \check{\bf r}' = 0 
                \Big\rangle 
\over   \Big\langle \check{\bf r}' = 0 \Big| 
                \exp\left\{ -\beta H_r \right\} 
        \exp\left\{ i H_r t \right\} \Big| \check{\bf r}' = 0 
                \Big\rangle 
} \,.
\label{normed}
\end{equation}
We should note that the real time dependence that appears here --- the
$t$ dependence --- makes a negligible contribution, and one can set
$t=0$ here. This time dependence only gives rise to energy
alterations on the order of particle thermal energies, which are
negligible compared to the energy release of the nuclear reaction. 
Using this notation and the results that we have thus far obtained,  
Eq.~(\ref{correct}) becomes, with the ``effective'' caveats noted
above still in place,
\begin{eqnarray}
&&  \Big\langle
        \langle {\bf r}_1'=0 = {\bf r}_2' \,,\, -i\beta | 
         U(\beta)
        | {\bf r}_1'= {\bf r} = {\bf r}_2' \,,\, t \rangle
        \Big\rangle_\beta
\nonumber\\
&&  \simeq 
        \langle {\bf r}_1'=0 = {\bf r}_2' , -i\beta | 
         {\bf r}_1'= {\bf r} = {\bf r}_2' , t \rangle
\Bigg\{
    1 + {1\over2} \int_0^\beta d\tau  d\tau'  
        \int { (d{\bf k}) \over (2\pi)^3 } G_\beta(k,\tau - \tau') 
                C(k,\tau - \tau')
\nonumber\\
&& \qquad \Big(\Big| \Bigg( e_1^2 \exp\left\{ i {\bf k} \cdot [
\check{\bf r}(\tau) - \check{\bf r}(\tau') ] M_2/M \right\}
        +  e_2^2 \exp\left\{- i {\bf k} \cdot [
\check{\bf r}(\tau) - \check{\bf r}(\tau') ] M_1/M \right\}
\nonumber\\
&& \qquad \qquad
        + 2 e_1 e_2 \exp\left\{ i {\bf k} \cdot \left[
{ M_2 \over M} \check{\bf r}(\tau) + 
        { M_1 \over M} \check{\bf r}(\tau') \right]  \right\}
                \Bigg)_+ \Big|\Big) \Bigg\} \,.
\label{tired}
\end{eqnarray}
To the order that we require, the potential correlation function
$G_\beta(k,\tau - \tau')$ is given by the ``sum of ring graphs"
approximation, as discussed in Appendix A. To get the leading terms
for dilute plasmas, a simplification can be made by subtracting the
value at $k=0$ of the function that multiplies $G_\beta(k,\tau -
\tau')$ in Eq.~(\ref{tired}), and adding it back as a separate
term. In the difference term we can replace the ``ring sum" for
$G_\beta(k,\tau - \tau')$ by the lowest order bubble. In this way we
simplify the integration in a manner that avoids a potential infrared
problem and captures all terms of order $e^3$ and $e^4$ in the final
result for the rate.  For the term in which the multiplying function
is evaluated at $k=0$, we retain the complete $G_\beta(k,\tau -
\tau')$ whose long-distance Debye screening removes the infrared
divergence, and we encounter an expression evaluated in
Eq.~(\ref{gulp}) of Appendix B,
\begin{equation}
\int_0^\beta d\tau d\tau' \int {(d{\bf k}) \over (2\pi)^3} 
G_\beta(k,\tau - \tau') = \beta \kappa_D +  8 
 \int_0^\beta d\tau d\tau' \int_0^\infty dk \, {1 \over k} 
{d \over dk}  \Pi^{(0)}(k,\tau - \tau') \,.
\label{infra}
\end{equation}
Here the first term involving the Debye wave number $\kappa_D$, which
is of order $e$ not $e^2$, isolates the long-distance contribution.
The second term involves the first-order polarization function
computed in Appendix A, whose dilute form reads
\begin{equation}
\Pi^{(0)}(k, \tau - \tau') \simeq {\sum}_s e_s^2 \,
\langle n_s \rangle^{(0)}_\beta C_s(k,\tau - \tau') \,,
\end{equation}
where
\begin{equation}
C_s(k,\tau - \tau') = \exp\left\{ - {  \beta k^2 \over 2M_s }
f(\tau - \tau') \right\} \,.
\end{equation}
Having removed the long-distance contribution, we may now make the
replacement 
\begin{equation}
G_\beta(k,\tau - \tau') \simeq \left( { 4\pi \over k^2} \right)^2 
\Pi^{(0)}(k, \tau - \tau') 
\end{equation}
in the remaining terms, which are of the form
\begin{equation}
G_\beta(k,\tau - \tau') \left[ C(k,\tau - \tau') e^{i {\bf k} \cdot
{\bf r} } - 1 \right] \,.
\end{equation} 
The time ordering involves the operators $ \check{\bf r}(\tau) $ and $
\check{\bf r}(\tau') $, each of whose components commute amongst
themselves. Hence the operators in the time-ordered product may be
treated like ordinary numbers in intermediate calculations. Thus we
now encounter a sum of the form
\begin{equation}
 {\sum}_s e_s^2 \langle n_s \rangle^{(0)}_\beta
\int_0^\beta d\tau d\tau' 
\end{equation}
of terms of the form
\begin{eqnarray}
\int { (d{\bf k}) \over (2\pi)^3 } \left( { 4\pi \over k^2} \right)^2
&& \exp\left\{ - { \beta  k^2 \over 2M_s }
f(\tau - \tau') \right\}
\nonumber\\
&&
e_a^2 \left[ \exp\left\{ - { \beta  k^2 \over 2M }
f(\tau - \tau')
\right\} e^{ i {\bf k} \cdot {\bf r} } - 1 \right] \,.
\label{terms}
\end{eqnarray}

To evaluate this expression, we represent
\begin{equation}
\left( { 1 \over k^2 } \right)^2 = \int_0^\infty s \, ds \,
 e^{- s k^2 } 
\end{equation}
and complete the square in the Gaussian wave number integrals
to evaluate them and write the expression (\ref{terms}) as
\begin{equation}
\sqrt{4\pi} e_a^2 \int_0^\infty sds \left[ (s+a_s)^{-3/2} \exp\left\{
- { {\bf r}^2 \over 4 (s+a_s) } \right\} - (s+b_s)^{-3/2} \right] \,,
\label{interr}
\end{equation}
where
\begin{equation}
a_s = { \beta \over 2 \bar M_s } f(\tau - \tau') \,,
\end{equation}
with $\bar M_s$ the reduced mass defined by
\begin{equation}
{ 1 \over \bar M_s } = { 1 \over M } + { 1 \over M_s } \,,
\end{equation}
and
\begin{equation}
b_s = { \beta \over 2 M_s } f(\tau - \tau') \,.
\end{equation}
The potential infrared divergence is reflected in the large $s$
behavior of the integrand in Eq.~(\ref{interr}) in which the separate
terms do not give convergent integrals. The cancellation of the
divergent pieces is exhibited by partial integration of the leading
terms which reduces the expression in Eq.~(\ref{interr}) to
\begin{equation}
- \sqrt{\pi} e_a^2 \int_0^\infty ds { 1 \over
(s+a_s)^{3/2} } \left[ 4a_s + { s \over s + a_s } {\bf r}^2 \right]
\exp\left\{ - { {\bf r}^2 \over 4 (s+a_s) } \right\} 
        + 8 e^2_a \sqrt{ \pi b_s} \,.
\label{interrr}
\end{equation}
It is easy to check that the last term here involving $\sqrt b_s$ gives
contributions that precisely cancel the second term on the right-hand
side of Eq.~(\ref{infra}). Omitting this term and making a simple
change of variables presents the result as
\begin{equation}
- 4 e_a^2 \sqrt{ \pi a_s} \int_0^1 { du \over \sqrt u} \, \left[ 1 +
{ (1-u) \over 4 a_s } {\bf r}^2 \right] \, 
\exp\left\{ - { u {\bf r}^2 \over 4 a_s } \right\} \,.
\end{equation} 

As is discussed in Appendix B, the contribution of the electrons in
the plasma should be separated out because their mass is so small
relative to the nuclear reacting particles and the nuclei in the
plasma. 
%On one hand, this small mass implies that the thermal electron
%distribution in the plasma may well not be dilute in the sense that it
%cannot be described by the Maxwell-Boltzmann distribution that we have
%been using, with the electrons needing the full Fermi-Dirac
%distribution. On the other hand, 
The relatively small electron mass
means that effectively $a_s \to \infty$ or that we may take ${\bf r}^2
= 0 $ for their contribution in Eq.~(\ref{interrr}). Thus, for the
electron contribution, the reacting particles may be taken to be at
the same spatial position, and the calculation of their effect is the
same as that for the number density calculation presented in Appendix
B, but for a particle of charge $ e_1 + e_2 $. 
%Using the results of
%this Appendix in addition to those that we have now secured, we arrive
%at
We evaluate the electronic part of the second term on the
right-hand-side of Eq.~(\ref{infra}) using the results of this
Appendix, which include the effects of Fermi-Dirac statistics
\footnote{Note that the Debye wave number which appears here should
be computed with the quantum-mechanical form (\ref{qdebye}) for the
electrons presented in Appendix A [(\ref{onedebye})] if the Fermi
statistics for these particles is important.}, to obtain 
\begin{eqnarray}
    \Big\langle
        \langle {\bf r}_1'=  0  = {\bf r}_2' && \,,\, -i\beta | 
         U(\beta)
        | {\bf r}_1'= {\bf r} = {\bf r}_2' \,,\, t \rangle
        \Big\rangle_\beta
\nonumber\\
&&  \simeq 
        \langle {\bf r}_1'=0 = {\bf r}_2' , -i\beta | 
         {\bf r}_1'= {\bf r} = {\bf r}_2' , t \rangle
\nonumber\\
&& \qquad
\Bigg\{
    1 + {1\over2} (e_1 + e_2)^2 \beta \kappa_D 
        - { 1 \over 8} (e_1 + e_2)^2 \beta 
        { 2 e^2 m_e \over e^{ -\beta \mu_e} + 1} - X \Bigg\} \,,
\label{greatest}
\end{eqnarray}
in which
\begin{eqnarray}
&& X = 
 \sum_{s \neq e} e_s^2 \langle n_s \rangle^{(0)}_\beta
        \int_0^\beta d\tau d\tau'
        \sqrt{4\pi a_s}  \int_0^1 { du \over \sqrt u }
\nonumber\\
&& \quad \Big(\Big| \Bigg( 
 e_1^2 \left[ 1 + { (1-u) \over 4 a_s }  
\left[ \check{\bf r}(\tau) - \check{\bf r}(\tau') \right]^2 (M_2/M)^2
\right]
\exp\left\{ - u { [
\check{\bf r}(\tau) - \check{\bf r}(\tau') ]^2 (M_2/M)^2
        \over 4 a_s } \right\}
\nonumber\\
&& \quad  + e_2^2 \left[ 1 + { (1-u) \over 4 a_s }
\left[ \check{\bf r}(\tau) - \check{\bf r}(\tau') \right]^2 (M_1/M)^2
\right]
\exp\left\{ - u { [
\check{\bf r}(\tau) - \check{\bf r}(\tau') ]^2 (M_1/M)^2
        \over 4 a_s } \right\}
\nonumber\\
&+&   
        2e_1 e_2 
 \left[ 1 + { (1-u) \over 4 a_s } 
\left[ { M_2 \over M} \check{\bf r}(\tau) + 
        { M_1 \over M} \check{\bf r}(\tau') \right]^2
        \right]
\exp\left\{ - u { 
\left[ { M_2 \over M} \check{\bf r}(\tau) + 
        { M_1 \over M} \check{\bf r}(\tau') \right]^2
        \over 4 a_s } \right\}
\Bigg)_+ \Big|\Big)  \,.
\nonumber\\
\left. \right.
\label{greater}
\end{eqnarray}

As we shall soon see, the third term inside of the curly bracket in
Eq.~(\ref{greatest}) is the most important correction, after the
Salpeter correction, for the case of a weakly coupled plasma. It
involves the quantum mechanics of the plasma in an essential way. This
fact leads to a dilemma in considering how to approach the regime of
lower temperature or higher density, a region in which the
perturbation expansion is not valid. The standard approach to this
domain is to use non-perturbative methods to treat the plasma
classically, and then to use the extracted effective potentials in the
tunneling calculation that determine the rates. But the fact that the
leading perturbative correction to the Salpeter result requires a
quantum treatment of the plasma argues that the quantum mechanics of
the plasma will be central to any meaningful calculation of the
strongly interacting case.

The correction $ X $ due the ionic component of the plasma is quite
small for a weakly coupled plasma, as a bound which we shall shortly
derive proves. But before doing this, it is worth giving more detail
to explain the remark of the previous paragraph that the character of
this correction for the weakly coupled case brings into question the
standard classical approach for dense plasmas. The matrix elements of
the position operator $\check {\bf r}(\tau)$ are controlled by the
magnitude of the classical turning point $r_{\rm max}$ discussed above
[Eq.~(\ref{turn})]. Thus the exponents in Eq.~(\ref{greater}) involve
the parametric ratio $ r_{\rm max}^2 / a_s $. In terms of the thermal
wave length
\begin{equation}
\bar \lambda_s^2 = { 2 \pi \beta \over \bar M_s } \,, 
\end{equation}
the denominator in this parametric ratio appears as
\begin{equation}
a_s = { \bar \lambda_s^2 \over 4\pi } f(\tau,\tau') \,.
\end{equation}
The thermal wave length $ \bar \lambda_s$ is a quantum length ---
in ordinary units, it is of order Planck's constant $\hbar$. 
At the center of a typical star like the sun, this quantum-mechanical
parametric 
ratio is about unity. In such cases, no small parameter appears in the
ionic correction, no further approximations may be performed, and
a full quantum-mechanical evaluation is required. It
must be emphasized that the {\em relevant ratio} $ r_{\rm max} / \bar
\lambda_s$ {\em entails the time-dependent, quantum-mechanical aspect
of the plasma}. {\em A static, classical treatment of the plasma for
such corrections is unphysical and gives incorrect results}. 

We now turn to bound the  correction $X$  and then evaluate this bound
approximately. The bound is easily obtained from the remark that the
quantum expectation values entailed in the correction $X$ have path
integral representations of the form 
\begin{equation}
  \Big\langle \check{\bf r}' = 0 \Big| 
                \exp\left\{ -\beta H_r \right\} \, 
\Big( F( \check{\bf r}(\tau) , \check{\bf r}(\tau') ) \Big)_+ 
\Big| \check{\bf r}' = 0 \Big\rangle 
=  \int [ d {\bf r} ] e^{-S} F( {\bf r}(\tau) , {\bf r}(\tau') )
\,,
\label{value}
\end{equation}
in which the paths start at $ {\bf r}(0) = 0$ and end at $ {\bf
r}(\beta) = 0$. The action functional $S$ is the classical action
continued to imaginary time. Since the action $S$ is real, the
measure $[ d {\bf r} ] \exp \{ - S \} $ is positive. Thus the
expectation value (\ref{value}) is bounded when the numerical
function $ F( {\bf r}(\tau) , {\bf r}(\tau') ) $ is
bounded. Denoting an exponent which appears in Eq.~(\ref{greater}) by
$u g$, we note that the bounded combinations
\begin{equation}
\left| ( 1 - u g) e^{-ug} \right| \le 1
\end{equation}
appears, giving the resulting parameter integral
\begin{equation}
\int_0^1 { du \over \sqrt{u} } = 2 \,.
\end{equation}
Following the evaluation of Eq.~(\ref{clever}) described in Appendix B, 
the imaginary time integrals for these terms involve
\begin{equation}
\int_0^\beta d\tau d\tau' \, \sqrt{ 4\pi a_s } = { \pi \over 8}
\beta^2 \bar \lambda_s \,.
\end{equation}
Thus, bounding the remaining terms with
\begin{equation}
\left| \int_0^1 { du \over \sqrt{u} } \, e^{- ug} \right|
\le \int_0^\infty { du \over \sqrt{u} } \, e^{-ug} = \sqrt{ \pi \over g}
\,,
\end{equation}
we obtain the bound 
\begin{eqnarray}
&& |X| \le 
 \sum_{s \neq e} \beta \kappa^2_{D,s}
                \Bigg\{ { 1 \over 16} \left( e_1 + e_2 \right)^2 
        \bar \lambda_s +  {1 \over 4 \beta^2} \int_0^\beta d\tau d\tau' 
\nonumber\\
&&   \Big(\Big| \Bigg( 
 \left( e_1^2 { M_2 \over M} + e_2^2 {M_1 \over M} \right) 
\left| \check{\bf r}(\tau) - \check{\bf r}(\tau') \right|
+  2 e_1 e_2 
\left| { M_2 \over M} \check{\bf r}(\tau) + 
        { M_1 \over M} \check{\bf r}(\tau') \right|
\Bigg)_+ \Big|\Big) \Bigg\} \,,
\label{geee}
\end{eqnarray}
where
\begin{equation}
\kappa_{D,s}^2 = 4\pi \beta e_s^2 \langle n_s \rangle_\beta^{(0)}
\end{equation}
is the contribution of the plasma species $s$ to the squared Debye
wave length.

We should note that although the first term in the curly brackets in
the bound (\ref{geee}) involves the quantum-mechanical nature of the
plasma since it involves the quantum-mechanical wavelength $\bar
\lambda_s$, the second set of terms refer only to the classical aspect
of the plasma, albeit as measured by quantum expectation values of the
reacting particles. This second set of terms is, in fact, just the
classical plasma limit of the ionic correction (\ref{greater}).  They
are obtained from the result (\ref{greater}) if the limit $\bar
\lambda_s \to 0$ is taken. This is the limit in which the plasma is
treated classically, the formal $\hbar \to 0$ limit, or the limit in
which the plasma particles are taken to have infinite mass (along with
the total mass of the reacting pair). This limit is equivalent to
taking $ a_s \to 0$. In the limit, the exponentials completely damp
out the $u$ integration except at the end point $u=0$. Hence, the $u$
integration may be extended to infinity. The terms which survive in
the limit yield precisely this second set of terms in
Eq.~(\ref{geee}). The connection of this limit with a computation which
starts out with a classical plasma is spelled out in some detail in
Appendix C.

To estimate the size of the remaining quantum expectation values in
Eq.~(\ref{geee}), we note, as has been previously mentioned, that the
basic quantum tunneling process is well described by the steepest
descent approximation to the path integral. Thus we may estimate the
size of these expectation values by replacing their quantum operators
by the classical solution to the Coulomb problem in imaginary time
which obeys the boundary conditions $ {\bf r}(0) = 0 = {\bf r}(\beta)
$.\footnote{The relative error to this classical evaluation is given
by the first quantum correction which is of order $ \Delta r^2 ( 2 /
r_{\rm max})^2 $, where $\Delta r^2$ is the average squared quantum
fluctuation about the classical path. As an examination of the
quadratic fluctuation correction to the path integral shows, this
quantum fluctuation is of the same order as that for a harmonic
oscillator with the same frequency, namely $ \Delta r^2 \simeq 
1 / (m \omega) = \beta / (2 \pi m) $. Using Eq's.~(\ref{turn}) and
(\ref{coul}), we find that $ \Delta r^2 ( 2 / r_{\rm max})^2 \simeq 1
/ \bar \eta $, which is negligible when the Coulomb parameter $\bar
\eta$ is large. At the center of the sun, $ \bar \eta \simeq 2 $, and
so our classical evaluation of the bound is a little rough.}
It is easy to verify that this solution to
\begin{equation}
{ d^2 {\bf r}(\tau) \over d \tau^2} = - { e_1 e_2 \over r(\tau)^3 }
        \, {\bf r}(\tau) \,,
\end{equation}
which gives the stationary path, has the parametric representation
\begin{equation}
\tau = { \beta \over 2\pi } ( \xi - \sin \xi ) \,,
\end{equation}
\begin{equation}
{\bf r}(\tau) = { 1 \over 2} r_{\rm max} \, \hat {\bf k} \, 
( 1 - \cos \xi ) \,,
\end{equation}
where $ 0 \le \xi \le 2\pi $, $r_{\rm max}$ is the classical turning
point given in Eq.~(\ref{turn}), and $\hat {\bf k}$ is an arbitrary
unit vector. The resulting integral
\begin{eqnarray}
I &=& \int_0^\beta d\tau d\tau' \, \left| {\bf r}(\tau) - {\bf
r}(\tau') \right| 
\nonumber\\
&=& \left( { \beta \over 2\pi } \right)^2 { 1 \over 2} \, r_{\rm max}
\int_0^{2\pi} d( \xi - \sin \xi ) \, d( \xi' - \sin \xi' ) \left| \cos
\xi - \cos \xi' \right| 
\end{eqnarray}
may be computed by partial integrations which lead to
\begin{equation}
{d \over dx } { d \over dx'} \left| x - x' \right| = -2 \delta ( x -
x' ) \,.
\end{equation}
Thus
\begin{equation}
I = { 8  \, r_{\rm max} \, \beta^2 \over 3 \pi^2} \,.
\end{equation}
The remaining integral is given by
\begin{eqnarray}
J &=& \int_0^\beta d\tau d\tau' \, \left| { M_2 \over M} {\bf r}(\tau) +
{ M_1 \over M} { \bf r}(\tau') \right|
\nonumber\\
&=& \beta { M_2 + M_1 \over M} {\beta \over 2\pi } { 1 \over 2}
r_{\rm max} \int_0^{2\pi} d\xi ( 1 - \cos \xi )^2
\nonumber\\
&=& {3 \over 4} r_{\rm max} \, \beta^2  \,.
\end{eqnarray}
Within this approximation we have 
\begin{equation}
 |X| \le 
 \sum_{s \neq e} \beta \kappa_{D,s}^2
        \Bigg\{ { 1 \over 16} \left( e_1 + e_2 \right)^2  \bar \lambda_s 
        +  r_{\rm max} \left[ { 2 \over 3\pi^2 }
\left( e_1^2 {M_2 \over M} + e_2^2 { M_1 \over M} \right) + { 3 \over
8 } e_1 e_2  \right] \Bigg\} \,. 
\label{bounder}
\end{equation}

We return to our result (\ref{greatest}) and omit the correction
due to the ionic component of the plasma. Using the Bohr radius
of the hydrogen atom $a_0 = 1 / e^2 m_e$ and noting that the terms
involving $e_1^2$ and $e_2^2$ just provide the corrected number
density -- chemical relation which is described in Appendix B, we find
that the plasma corrected rate is given by
\begin{equation}
\Gamma_P = \left\{ 1 + \beta e_1 e_2 \left[ \kappa_D - { 1 \over
2 a_0 } { 1 \over e^{ - \beta \mu_e } + 1 } \right] 
\right\} \, \Gamma_C \,,
\label{endresult}
\end{equation}
where $ \Gamma_C$ is the rate in the absence of plasma corrections
given in Eq.~(\ref{coulrate}). In writing this result we have
neglected the quantum effects described by the correction $X$ given in
Eq.~(\ref{greater}). For the example given by the center of the sun,
$r_{\rm max} / \lambda \approx 1/2$, and $ \kappa_D \lambda \approx 2
\times 10^{-2}$. Taking $e_1 = e_2$, we find that $ |X| < \beta e_1
e_2 \kappa_D \times 1 \times 10^{-2} $. That is, the quantum
correction $X$ is less than 1\% of the basic Salpeter correction. At
the center of the sun, the electron correction that has been kept in
the result (\ref{endresult}) above, the second term in the square
brackets, involves $ 1/ 2 a_0 \approx \kappa_D / 4$ and $ \exp\{ \beta
\mu_e \} \approx 0.3 $. Thus it forms a correction of about 10\%
relative to the basic Salpeter correction. Moreover, it should be kept
in mind that with $ \exp\{ \beta \mu_e \} \approx 0.3$, the correct
quantum expression (\ref{qdebye}) for the electron contribution to
the Debye wave number is reduced by about 10\% relative to its
classical value (\ref{cdebye}).

Since the last, electronic correction which appears in
Eq.~(\ref{endresult}),
\begin{equation}
 \left[ \kappa_D - { 1 \over
2 a_0 } { 1 \over e^{ - \beta \mu_e } + 1 } \right]  \,,
\end{equation}
is the leading correction to the basic Salpeter
result, we should describe its nature in a little more detail. In our
formal counting of orders in the electric charge $e$, the Debye wave
number $\kappa_D$ is of order $e$, while the new correction is of
order $e^2$. The inverse Bohr radius $1/a_0$ is proportional to one
power of Planck's constant $\hbar$, and so the new correction
explicitly involves the quantum-mechanical character of the
plasma. The final factor with the electron's chemical potential
represents the effect of Fermi-Dirac statistics. When the electrons
are sufficiently dilute ($ \exp\{ \beta \mu_e \} \ll 1 $) 
so that the Boltzmann limit can be used, the new correction may be
written as
\begin{equation}
 \left[ \kappa_D - { 1 \over 8 } \kappa^2_{D,e} \lambda_e \right] \,. 
\end{equation}
Here $\kappa^2_{D,e}$ is the electronic contribution to the squared
Debye wave length, and $\lambda_e$ is the electron's thermal wave
length defined long ago in Eq.~(\ref{donne}). The thermal wave length
is again proportional to $\hbar$, and we again see that the new
correction is a first quantum correction.

\section{Conclusions}

We have developed a concise expression [Eq.~(\ref{great})] for the
combined effects of the surrounding plasma and the Coulomb barriers on
reaction rates in a completely ionized plasma. The expression is
applicable to any case in which the actual reaction takes place at
very small particle separations, and in which the reaction rate is
slow, in a well defined sense. In general, there can be significant
Coulomb and plasma effects in both initial and final states, and these
effects are entangled in the results. For the cases of importance in
solar physics, in which the release of energy in the reaction is very
large compared to thermal energies, the correction reduces to one
which involves a factor [Eq.~(\ref{probb})] that is a generalization
of the probability that one fusing ion is at the position of the
other. This generalization takes account of the effects of the
center-of-mass motion of the initial fusing nuclei through the plasma. 

We addressed the calculation of this quantity in an approach that is
based on perturbation theory in the couplings of the plasma particles
to the two (distinguished) particles that will undergo fusion. The
Coulomb force between the fusing particles is treated
nonperturbatively through use of the action-minimizing imaginary time
paths for tunneling problems; alternatively our expressions could be
expanded in Coulomb wave functions. The interactions among the plasma
particles were treated in one-loop order. Expressing the result for
the rate as a correction factor times the result in the absence of the
plasma, our calculation captures all terms in the correction factor of
order $e^3$ and $e^4$ .

The results consisted of a recapitulation of the standard Salpeter
results for the effects of the electron and ionic plasma particles,
plus a single new analytical term coming from the quantum nature of
the electron plasma, plus a rigorous bound on the remaining terms of
our expression, which turn out to make negligible contributions under
solar conditions. We have two conclusions that are significant to
calculations of solar processes:

1) The quantum mechanical term in the dynamics of the electron plasma
gives a reduction in the fusion rate of about 10 percent of the
Salpeter enhancement term, under the conditions that prevail in the
core of the sun.

2) There is no ``dynamical screening" modification of the Salpeter
result, to leading order in the dimensional parameters.  There is no
dependence on the masses of the plasma ions of the terms that are
appreciable under solar conditions.  In consequence, the fusion rates
are somewhat greater than those calculated by Carrero {\it et al}.\
(1988). In Appendix D we demonstrate both how such terms arise in a
real time approach and how they are cancelled.\footnote{ As of the
time of the writing of this paper it seems likely that the most
critical calculation of rates in the sun, for the purpose of
understanding the solar neutrino signal, will be the comparison of the
$ p + ^7{\rm Be} \to \gamma + ^8{\rm B}$ rate with that of the rate of
electron capture on $^7{\rm Be}$ [Bahcall (1995), Hata and Langacker
(1995)]. Modification (1) gives about a 2\% decrease and modification
(2) about a 2\% increase for the $ p + ^7{\rm Be}$ reaction. Thus our
results have little impact on the fusion side of the
comparison. However, it would be interesting to evaluate the plasma
effects on the electron capture process using the methods of this
paper. }

The nature of the remaining small correction terms in our results, and
the formal structure of the results as well, argue against an approach
to the strongly coupled problem in which numerical results for the
classical plasma are turned into effective two body potentials,
followed by the quantum tunneling calculation to obtain rates. In the
weakly coupled domain, however, the result of our systematic
perturbation theory provides support for the way in which screening
calculations have been used in the literature, subject to the
modifications mentioned above. In particular, the interparticle
spacing does not have to be much smaller than the Debye length, even
though this assumption goes into the derivation of the classical Debye
potential.\footnote{We thus take issue with the remarks of Dar and
Shaviv (1996), who question the applicability of the standard
screening lore under solar conditions, based on the fact that this
condition on the interparticle spacing is not satisfied.}  We
reemphasize the fact that our treatment nowhere used this potential.

The results of section 2 are general, and could be the first step in
developing approximations for the strongly coupled case. The
perturbation development in section 3 is applicable only to the weak
screening domain. In view of the smallness of the  $e^4$ terms that we
found, this fact seems to present no problems under solar conditions,
but will probably not be useful for the interiors of much smaller main
sequence stars or more highly evolved stars.

\begin{center}

ACKNOWLEDGMENTS

\end{center}

This project started at the Aspen Center for Physics. 
One of the authors (LSB) has had fruitful discussions on this topic
with L.~G. Yaffe. He would also like to thank R.~S. Steinke for
checking the manuscript. This work was supported, in part, by the
U.S. Department of Energy under Grant No. DE-FG03-96ER40956.

\appendix

\section{Thermodynamic Field Theory Review}

We review and collect here some results of multiparticle,
non-relativistic field theory at finite temperature that are used in
our work.\footnote{The review here is close in spirit to discussion in
Chapter 2 of Brown (1992), especially Problems 3 and 4.}

The dependence in imaginary time of an arbitrary operator is given by
\begin{equation}
X(\tau) = e^{H\tau} X e^{-H\tau} \,.
\label{time}
\end{equation}
The thermal average of two operators is defined by
\begin{equation}
\langle A(\tau) B(\tau') \rangle_\beta = Z^{-1} \, {\rm Tr} 
\exp\left\{ - \beta \left[ H - {\sum}_a \mu_a N_a \right] \right\}
        A(\tau) \, B(\tau') \,,
\end{equation}
where
\begin{equation}
Z = {\rm Tr} 
\exp\left\{ - \beta \left[ H - {\sum}_a \mu_a N_a \right] \right\}
\end{equation}
is the grand canonical partition function. Thus, if the operators
commute with the particle number operators $N_a$, there is the cyclic
symmetry
\begin{equation}
\langle A(\tau) B(\tau') \rangle_\beta = \langle B(\tau') A(\tau - \beta)
\rangle_\beta \,,
\label{cyclic}
\end{equation}
which follows from the cyclic symmetry of the trace and the time
dependence given in Eq.~(\ref{time}). 
Accordingly, the Green's function --- the ordered product in imaginary
time --- of two operators,
\begin{equation}
G(\tau - \tau') = \left\langle \Big( A(\tau) B(\tau') \Big)_+ 
\right\rangle_\beta \,,
\end{equation}
is periodic in imaginary time with period $\beta$, and it may be
expanded in a Fourier series,
\begin{equation}
G(\tau - \tau') = {1 \over \beta} \, {\sum}_n 
e^{-i\omega_n (\tau - \tau')}  \, g(i\omega_n) \,,
\end{equation}
where
\begin{equation}
\omega_n =  2\pi n / \beta \,.
\end{equation}

Going over to real time $(\tau \to it)$, with
\begin{equation}
X(t) = e^{i H t} X e^{-i H t} \,,
\end{equation}
we write the thermal average of the commutator as the Fourier integral
\begin{equation}
\langle \, [ A(t) , B(t') ] \, \rangle_\beta = \int_{-\infty}^\infty
 {d\omega \over 2\pi } \, c(\omega) \,  e^{-i \omega (t - t') } \,.
\end{equation}
Using the cyclic symmetry (\ref{cyclic}), a short calculation shows
that, in the imaginary time interval $ 0 < \tau < \beta$,
\begin{equation}
\langle A(\tau) B(0) \rangle_\beta = \int_{-\infty}^\infty {d\omega
\over 2\pi} {c(\omega) \over 1 - e^{-\beta \omega} } e^{-\omega \tau}
\,.
\end{equation}
Therefore,
\begin{eqnarray}
g(i\omega_n) &=& \int_0^\beta d\tau \langle A(\tau) B(0)
\rangle_\beta \, e^{i\omega_n \tau} 
\nonumber\\
&=& \int_{-\infty}^\infty {d\omega \over 2\pi } { c(\omega) \over
\omega - i\omega_n } \,.
\label{disp}
\end{eqnarray}

To relate these results to the charge-density correlations that we
need for our work, we note that first-order perturbation theory
shows that the introduction of a small external potential $\delta
\phi_{\rm ext}({\bf r},t) $ induces an average charge density given by
\begin{equation}
\langle \delta \rho({\bf r},t) \rangle_\beta = \int (d{\bf r}')
        \int dt' R( {\bf r} - {\bf r}' , t - t') 
        \, \delta \phi_{\rm ext}({\bf r}',t') \,,
\end{equation}
where
\begin{eqnarray}
R({\bf r} - {\bf r}' , t - t') &=& -i \langle \, [ \rho({\bf r},t) ,
\rho({\bf r}' , t') ] \, \rangle_\beta \, \theta (t-t')
\nonumber\\
&=& \int { (d {\bf k}) \over (2\pi)^3} \int {d\omega \over 2\pi}
  \, r( k, \omega) \, e^{ i {\bf k} \cdot ( {\bf r} - {\bf r}')
        -i \omega (t-t') } \,.
\label{retcomm}
\end{eqnarray}
The similar Fourier transform of the commutator function itself,
without the factor $-i$ and without the step function $\theta(t-t')$,
defines a weight which we shall denote by $c(k,\omega)$. Since this
commutator function is odd under the interchange $t \leftrightarrow
t'$,  $c(k,-\omega) = - c(k,\omega)$ is odd in $\omega$. 
Since complex conjugation of
the expectation value the commutator of the Hermitian fields $\rho$ is
equivalent to this interchange, which is compensated by the complex
conjugation of $\exp\{ -i \omega (t-t') \}$, $c(k,\omega)^* =
c(k,\omega)$ is a real function.  The relationship between the Fourier
transform of the retarded commutator function which appears in
Eq.~(\ref{retcomm}) and the Fourier transform $ c(k,\omega)$ of the
commutator function itself is given by the dispersion relation
\begin{equation}
r(k,\omega) = \int_{-\infty}^\infty {d\omega' \over 2\pi}
{ c(k,\omega') \over \omega - \omega' -i\epsilon} \,,
\end{equation}
as one can easily prove by directly Fourier transforming over only
positive time differences the Fourier representation of the commutator
function. In view of the general relation (\ref{disp}) between the
commutator and Green's functions, we see that the charge-density
correlator in imaginary time, 
\begin{equation}
\left\langle \Big( \rho({\bf r},\tau) \rho({\bf r}',\tau') \Big)_+  
\right\rangle_\beta = {1 \over \beta} \, {\sum}_n e^{-i\omega_n (\tau -
\tau')} \int { (d {\bf k}) \over (2\pi)^3 } 
\, g(k , i\omega_n) 
e^{i {\bf k} \cdot ( {\bf r} - {\bf r}' ) } \,,
\end{equation}
has a weight given by the analytic continuation
\begin{equation}
g(k, i\omega_n ) = - r( k , i \omega_n ) \,.
\end{equation}
The induced charge -- perturbing potential relation in
Fourier space,
\begin{eqnarray}
\langle \delta \rho({\bf k},\omega) \rangle_\beta
        &=& r(k,\omega) \, \delta \phi_{\rm ext}({\bf k},\omega) 
\nonumber\\
&=& \left[ { 1 \over \epsilon(k,\omega) } -1 \right]
{ k^2 \over 4\pi} \, \delta \phi_{\rm ext}({\bf k},\omega) \,,
\end{eqnarray}
defines the dielectric function $ \epsilon(k,\omega)$. 
Hence,
\begin{equation}
g(k,i\omega_n) = { k^2 \over 4\pi } \left[ 1 - { 1 \over
\epsilon(k, i\omega_n ) } \right] \,.
\end{equation}
It is worthwhile noting that, reverting to ordinary units and
restoring Planck's constant $\hbar$, $ \omega_n = 2\pi n /\hbar
\beta$. Hence in the classical limit $ \hbar \to 0$, which is
equivalent to the high temperature limit $\beta \to 0$, $
\omega_n \to \infty$ if $ n \neq 0$ and only the $ n=0 $ term
contributes to the  Fourier sum defining the correlation
function. Thus in this limit, the correlation function becomes
independent of the imaginary times, with
\begin{equation}
\langle \rho({\bf r}) \rho({\bf r}') \rangle_\beta \simeq
{ 1 \over \beta } \int { (d {\bf k}) \over (2\pi)^3 } 
{k^2 \over 4\pi } 
\left[ 1 - { 1 \over \epsilon(k,0) } \right] 
e^{i {\bf k} \cdot ( {\bf r} - {\bf r}' ) } \,.
\label{rhoclass}
\end{equation} 

The dielectric function is related to the irreducible polarization
function by
\begin{equation}
\epsilon(k,\omega) = 1 + { 4\pi \over k^2 }
        \Pi(k,\omega) \,,
\end{equation}
which gives
\begin{equation}
g(k,i\omega_n) = { \Pi(k,i\omega_n) \over 1 + 
        { 4\pi \over k^2 } \Pi(k,i\omega_n) } \,.
\end{equation}
The ``sum of ring graphs'' approximation is obtained by approximating
the exact polarization function which appears here by its
lowest-order, one-loop form $\Pi^{(0)}$. This function is obtained by
neglecting the denominator correction above and setting
\begin{equation}
\Pi^{(0)}({\bf r} - {\bf r}' , \tau - \tau' ) =
\left\langle \Big( \rho({\bf r},\tau) \rho({\bf r}',\tau') \Big)_+
\right\rangle^{(0)}_\beta \,,
\end{equation}
where the superscript on the expectation value indicates that it is
computed for free fields. The charge density operator which appears
here has the field expression
\begin{equation}
\rho({\bf r},\tau) = {\sum}_s e_s \, \psi^\dagger_s({\bf r},\tau) 
        \psi_s({\bf r},\tau) \,,
\end{equation}
where the sum runs over all the particle species with charge $e_s$ (and
over all spin components). In the free field approximation,
\begin{equation}
\left\langle \Big( \rho({\bf r},\tau) \rho({\bf r}',\tau') \Big)_+
\right\rangle^{(0)}_\beta = {\sum}_s e_s^2 
\left\langle \Big( \psi_s({\bf r},\tau) \psi^\dagger_s({\bf r}',\tau')
\Big)_+ \right\rangle^{(0)}_\beta 
\left\langle \Big( \psi^\dagger_s({\bf r},\tau) \psi_s({\bf r}',\tau')
\Big)_+ \right\rangle^{(0)}_\beta \,,
\end{equation}
since the total charge vanishes and with it the expectation value of a
single charge operator. 

The field expectation values that appear here
are obtained from the generic forms
\begin{equation}
\left\langle \psi({\bf r},\tau) \, \psi^\dagger({\bf r}',\tau') 
 \right\rangle^{(0)}_\beta = \int { (d{\bf p}) \over (2\pi)^3 } 
\left[ 1 \pm n({\bf p}) \right] e^{i {\bf p} \cdot ( {\bf r} - {\bf
r}' ) } e^{- E({\bf p}) (\tau - \tau') } \,,
\label{freer}
\end{equation}
and
\begin{equation}
\left\langle  \psi^\dagger({\bf r}',\tau') \, \psi({\bf r},\tau) 
 \right\rangle^{(0)}_\beta = \int { (d{\bf p}) \over (2\pi)^3 } 
 n({\bf p})  e^{i {\bf p} \cdot ( {\bf r} - {\bf
r}' ) } e^{- E({\bf p}) (\tau - \tau') } \,,
\label{freel}
\end{equation}
where the upper $+$ sign is for Bosons, the lower $-$ sign is for
Fermions, 
\begin{equation}
E({\bf p}) = {\bf p}^2 / 2M 
\end{equation}
is the kinetic energy of the particle with mass $M$, and
\begin{equation}
n({\bf p}) = { 1 \over e^{-\beta [ E({\bf p}) - \mu] } \mp 1 }
\end{equation}
is the momentum distribution for free Bosons or Fermions.\footnote{It
is easy to confirm the validity of the results (\ref{freer}) and
(\ref{freel}): They obey the free field equations of motion, the
equal-time commutation or anticommutation relations, and they obey 
the cyclic boundary condition (\ref{cyclic}) generalized to include
the number operator and chemical potential in an effective
Hamiltonian. These conditions fix the results uniquely.} Using the
results (\ref{freer}) and (\ref{freel}), we have 
\begin{eqnarray}
\Pi^{(0)}({\bf r} - {\bf r}' , \tau - \tau' ) &=& {\sum}_s e^2_s
\int { (d{\bf k}) \over (2\pi)^3 } e^{i{\bf k} \cdot ( {\bf r} - {\bf
r}' ) } \int { (d{\bf p}) \over (2\pi)^3 } e^{- [ E_s({\bf p+k}) - 
E_s({\bf p}) ] (\tau - \tau') }
\nonumber\\
&& \qquad \qquad \left\{
\begin{array}{c}
n_s({\bf p}) [ 1 \pm n_s({\bf p+k}) ] \,, \qquad \tau > \tau' \,,
\\
n_s({\bf p+k}) [ 1 \pm n_s({\bf p}) ] \,, \qquad \tau' > \tau \,.
\end{array}
\right.
\label{pi}
\end{eqnarray}
The Fourier transform of this expression yields
\begin{equation}
\Pi^{(0)}(k, i\omega_n) = {\sum}_s e_s^2 
\int {(d{\bf p}) \over (2\pi)^3} \,
{n_s({\bf p+k}) - n_s({\bf p}) \over i\omega_n - [ E_s({\bf p+k}) - 
E_s({\bf p}) ] } \,.
\end{equation}

The long-distance behavior of the polarization is controlled by the
Debye wave number
\begin{equation}
\kappa_D^2 = 4\pi \, \Pi(0,0) \,.
\end{equation}
In the one-loop approximation,
\begin{eqnarray}
\kappa_D^2 & \simeq & - 4\pi \, {\sum}_s e_s^2 
\int {(d{\bf p}) \over (2\pi)^3}
{\partial n_s({\bf p}) \over \partial E_s({\bf p}) } 
\nonumber\\
&=&  4\pi \, {\sum}_s e_s^2 \, {\partial \over \partial \mu_s }
\int {(d{\bf p}) \over (2\pi)^3} n_s({\bf p}) 
= 4\pi \, {\sum}_s e_s^2 \, {\partial \over \partial \mu_s } 
\, \langle n_s \rangle^{(0)}_\beta \,.
\label{onedebye}
\end{eqnarray}
In the dilute gas limit, the Bose and Fermi distributions can be
replaced by the Maxwell distribution
\begin{equation}
n({\bf p}) \simeq e^{-\beta [ E({\bf p}) - \mu ]} \,,
\label{max}
\end{equation}
and in this limit
\begin{equation}
\kappa_D^2 = 4\pi \beta \, {\sum}_s e_s^2 
\, \langle n_s \rangle^{(0)}_\beta \,.
\end{equation}

The Debye wave number is essentially a classical quantity, and in the
dilute approximation the zero-frequency classical dielectric function
is given by
\begin{equation}
\epsilon(k,0) = 1 + { \kappa_D^2 \over k^2} \,.
\end{equation}
Placing this in the classical limit (\ref{rhoclass}) of the charge
density correlation function gives
\begin{equation}
\langle \rho({\bf r}) \rho({\bf r}') \rangle_\beta \simeq
{ 1 \over \beta } \int { (d {\bf k}) \over (2\pi)^3 } 
{k^2 \over 4\pi } 
{ \kappa_D^2 \over k^2 + \kappa_D^2 } 
e^{i {\bf k} \cdot ( {\bf r} - {\bf r}' ) } \,.
\end{equation} 
The corresponding potential correlation function is thus
\begin{equation}
\langle \phi({\bf r}) \phi({\bf r}') \rangle_\beta \simeq
{ 1 \over \beta } \int { (d {\bf k}) \over (2\pi)^3 } 
{4\pi \over k^2 } 
{ \kappa_D^2 \over k^2 + \kappa_D^2 } 
e^{i {\bf k} \cdot ( {\bf r} - {\bf r}' ) } \,.
\end{equation} 
The partial fraction decomposition
\begin{equation}
{1 \over k^2 } 
{ \kappa_D^2 \over k^2 + \kappa_D^2 } =
{1 \over k^2 } - 
{ 1 \over k^2 + \kappa_D^2 } 
\end{equation}
expresses the potential correlation function as the difference of the
Fourier representations of the Coulomb and Yukawa potentials, and so
\begin{equation}
\langle \phi({\bf r}) \phi({\bf r}') \rangle_\beta \simeq
{ 1 \over \beta |{\bf r} - {\bf r}' |} \left[ 1 - \exp\{ - \kappa_D
|{\bf r} - {\bf r}' | \} \right] \,.
\label{phiclass}
\end{equation} 

In the dilute gas limit, $ n({\bf p}) \ll 1$ and the Maxwell
distribution (\ref{max}) applies. Thus, in this limit, and for $ \tau
> \tau'$, the result (\ref{pi}) reduces to
\begin{equation}
\Pi^{(0)}(k, \tau - \tau') \simeq {\sum}_s e_s^2 \, e^{\beta \mu_s}
\int { (d{\bf p}) \over (2\pi)^3 } e^{-[ E_s({\bf p+k}) -E_s({\bf p})
](\tau - \tau') } e^{ - \beta E_s({\bf p}) } \,.
\end{equation}
Since the energies that appear in the exponentials are quadratic in
the momentum ${\bf p}$, the momentum integral is Gaussian which can be
evaluated by completing the square. The same considerations apply for
the other time order $ \tau' > \tau$, and one finds that for a dilute
gas
\begin{equation}
\Pi^{(0)}(k, \tau - \tau') \simeq {\sum}_s e_s^2 \,
\langle n_s \rangle^{(0)}_\beta C_s(k,\tau - \tau') \,,
\label{mist}
\end{equation}
where
\begin{equation}
C_s(k,\tau - \tau') = \exp\left\{ - { k^2 \over 2M_s }
| \tau - \tau' | \Big[ 1 - | \tau - \tau' | / \beta \, \Big] 
\right\} \,.
\label{pitwo}
\end{equation}

\section{Density Calculated by Reaction Rate Method}

The method employed in the text to compute the nuclear reaction rate
in a plasma may also be used to compute the average density of one of
the species of the particles in the plasma. This we shall do here so
as to illustrate the method in a simpler context and also to obtain
some results that are needed for the work of the text.\footnote{A
standard treatment of some of the results that we obtain here appears,
for example, in Fetter and Walecka (1971), Sec. 30.}

Using the techniques of the text, it is easy to see that the average
density of a particle with chemical potential $\mu$ may be expressed
as
\begin{equation}
\langle n \rangle_\beta = e^{\beta\mu} \langle {\bf r}' = 0, -i\beta |
\langle \, U(\beta) \, \rangle_\beta | {\bf r}' = 0 \rangle \,,
\end{equation}
where the overall matrix element is a single-particle matrix element
of the particle in question and now
\begin{equation}
U(\beta) = \left( \exp\left\{ - \int_0^\beta d\tau q
\phi({\bf r}(\tau),\tau ) \right\} \right)_+ \,,
\end{equation}
with $q$ the charge of the particle and ${\bf r}(\tau)$ the particle's
coordinate operator which undergoes free motion in the imaginary time
$\tau$. The only approximation involved here is the requirement that
the particle density $\langle n \rangle_\beta$ being computed is
dilute. The other components in the plasma may be dense. 

To illustrate
the method, we shall, however, compute the case in which all
the particles in the plasma are dilute so that the exponential in
$U(\beta)$ may be expanded with only the first non-vanishing
correction retained,
\begin{equation}
\langle U(\beta) \rangle_\beta \simeq 1 + { q^2 \over 2 } \int_0^\beta
d\tau d\tau'  \left\langle \Big( \phi({\bf r}(\tau),\tau) \,
\phi({\bf r}(\tau),\tau) \Big)_+ \right\rangle_\beta \,.
\end{equation}
Here
\begin{equation}
\left\langle \Big( \phi({\bf r},\tau) \,
\phi({\bf r}',\tau) \Big)_+ \right\rangle_\beta  = 
\int { (d{\bf k}) \over (2\pi)^3 } e^{ i {\bf k} \cdot 
( {\bf r} - {\bf r}' ) } G_\beta(k, \tau - \tau') \,,
\end{equation}
with, as is appropriate for the dilute case,
\begin{equation}
G_\beta( k, \tau - \tau') = { 1 \over \beta } \, {\sum}_n e^{ - i
\omega_n (\tau - \tau') } \left( { 4\pi \over k^2 } \right)^2
{ \Pi^{(0)}( k, i\omega_n) \over 1 + {4\pi \over k^2} 
\Pi^{(0)}( k,i\omega_n) } \,,
\end{equation}
where the lowest-order polarization function $\Pi^{(0)}$ has been
calculated in Appendix A.  The work in the text [{\it c.f.}
Eq.'s~(\ref{centerb}) and (\ref{centerc})] tells us that, with $M$ the
mass of the particle in question,
\begin{equation}
\langle {\bf r}' = 0 , -i\beta | {\bf r}' = 0 \rangle =  \left( { 
M \over 2\pi \beta} \right)^{3/2} \,,
\end{equation}
and
\begin{equation}
\langle {\bf r}' = 0 , -i\beta | \left( e^{ i{\bf k} \cdot [ {\bf
r}(\tau) - {\bf r}(\tau') ]} \right)_+ |
{\bf r}' = 0 \rangle =  \left( { 
M \over 2\pi \beta} \right)^{3/2} C(k, \tau - \tau') \,.
\end{equation}
Here
\begin{equation}
C(k, \tau - \tau') = \exp\left\{ - { k^2 \over 2M } | \tau
- \tau' | \Big[ 1 - |\tau - \tau'| / \beta \Big] \right\}
\label{cfn}
\end{equation}
is periodic in the imaginary times $\tau$ and $\tau'$ in the interval
$0\,,\beta$ as is the function $G_\beta(k, \tau - \tau')$. Since 
two such periodic functions may be expressed as Fourier series, the
double time integral reduces to a single integral,
\begin{equation}
\int_0^\beta d\tau d\tau' f(\tau - \tau') g(\tau - \tau') =
        \beta \int_0^\beta d\tau f(\tau) g(\tau) \,,
\end{equation}
and we may write our result as
\begin{equation}
\langle n \rangle_\beta \simeq e^{\beta\mu} 
\, \left( { M \over 2\pi \beta } \right)^{3/2}
 \left\{ 1 + \beta {q^2 \over 2} 
\int_0^\beta d\tau \int { (d{\bf k}) \over (2\pi)^3} C(k,\tau) 
G_\beta(k,\tau) \right\} \,.
\end{equation}

To evaluate the correction which appears here, we set
\begin{equation}
C(k,\tau) = 1 + [ C(k,\tau) - 1 ] \,,
\end{equation}
and note that $[C(k,\tau) - 1 ]$ vanishes as $k^2$ for
small $k$. Thus the long-ranged contribution of the potential
correlation function is reduced in this second term, it is no longer
singular at small $k$, and the potential correlation function may
be replaced by its zeroth-order value. On the other hand, the
replacement of $C(k,\tau)$ by unity just picks out the $n=0$
Fourier mode of the potential correlation function. Therefore, to
within an accuracy of order $e^2$,
\begin{eqnarray}
\int_0^\beta d\tau \int {(d{\bf k}) \over (2\pi)^3} C(k,\tau) 
& G_\beta & (k,\tau)  =  \int {(d{\bf k}) \over (2\pi)^3} 
\left( {4\pi \over k^2 } \right) 
{ 4\pi \Pi^{(0)}(k,0) \over k^2 + 4\pi \Pi^{(0)}(k,0) } 
\nonumber\\
 &+& \int_0^\beta d\tau \int {(d{\bf k}) \over (2\pi)^3} 
\Big[ C(k,\tau) - 1 \Big] \left( {4\pi \over k^2 } \right)^2 
\Pi^{(0)}(k,\tau) \,.
\label{nice}
\end{eqnarray}  

Recalling that $4\pi \Pi(0,0) = \kappa^2_D$ and noting that only
potential singular behavior of the integration at small $k$ can give a
contribution that is not of order $e^2$, we see that to this order we
may write
\begin{eqnarray}
\int {(d{\bf k}) \over (2\pi)^3} 
\left( {4\pi \over k^2 } \right) 
{ 4\pi \Pi^{(0)}(k,0) \over k^2 + 4\pi \Pi^{(0)}(k,0) } & = & 
\int {(d{\bf k}) \over (2\pi)^3} 
\left( {4\pi \over k^2 } \right) 
{ \kappa_D^2 \over k^2 + \kappa_D^2 } 
\nonumber\\
 &+&
\int {(d{\bf k}) \over (2\pi)^3} 
\left( {4\pi \over k^2 } \right)^2 
\Big[ \Pi^{(0)}(k,0) - \Pi^{(0)}(0,0) \Big] \,.
\end{eqnarray}
Here
\begin{equation}
\int {(d{\bf k}) \over (2\pi)^3} 
\left( {4\pi \over k^2 } \right) 
{ \kappa_D^2 \over k^2 + \kappa_D^2 } = \kappa_D \,,
\end{equation}
while
\begin{eqnarray}
\int {(d{\bf k}) \over (2\pi)^3} 
\left( {4\pi \over k^2 } \right)^2
\Big[ \Pi^{(0)}(k,0) - \Pi^{(0)}(0,0) \Big] &=&
- 8 \int_0^\infty dk \Big[ \Pi^{(0)}(k,0) - \Pi^{(0)}(0,0) \Big] 
{d \over dk} {1 \over k} 
\nonumber\\
&=&  8 \int_0^\beta d\tau \int_0^\infty dk \, {1 \over k} 
{d \over dk}  \Pi^{(0)}(k,\tau) \,,
\label{parts}
\end{eqnarray}
where we have used polar coordinates, integrated by parts, and
re-expressed the zero frequency component of the polarization function
as a integral over imaginary time. The results that we have just
obtained may be restated in the form used in Section 3.2 of the text, 
\begin{equation}
\int_0^\beta d\tau \int {(d{\bf k}) \over (2\pi)^3} 
G_\beta(k,\tau) = \kappa_D +  8 
 \int_0^\beta d\tau  \int_0^\infty dk \, {1 \over k} 
{d \over dk}  \Pi^{(0)}(k,\tau ) \,,
\label{gulp}
\end{equation}
which we reiterate is accurate to terms including order $e^2$.  

For the second set of terms in Eq.~(\ref{nice}), we again integrate by
parts as was done in Eq.~(\ref{parts}) to obtain
\begin{eqnarray}
\int_0^\beta d\tau && \int {(d{\bf k}) \over (2\pi)^3} 
\Big[ C(k,\tau) - 1 \Big] \left( {4\pi \over k^2 } \right)^2 
\Pi^{(0)}(k,\tau)  = 
\nonumber\\
&&  8 \int_0^\beta d\tau 
\int_0^\infty dk {1 \over k} \Big\{
 {d \over dk} \Big[ C(k,\tau) \Pi^{(0)}(k,\tau) \Big] 
 - {d \over dk} \Pi^{(0)}(k,\tau) \Big\} \,.
\end{eqnarray}
The second term in the integrand on the right-hand side of this
equation just cancels the contribution in Eq.~(\ref{parts}).
For the first term in the integrand, we recall the results
(\ref{mist}) and (\ref{pitwo}) from Appendix A which give the dilute
gas form
\begin{equation}
\Pi^{(0)}(k, \tau ) = {\sum}_s e_s^2 \,
\langle n_s \rangle^{(0)}_\beta 
 \exp\left\{ - { k^2 \over 2M_s }
 \, \tau \Big[ 1 -  \tau  / \beta \, \Big] \right\} \,.
\end{equation}
The exponential here is of the same form as that in the definition
(\ref{cfn}) of $C(k,\tau)$, and the two combine to form a single such
exponential which involves the reduced mass
\begin{equation}
{ 1 \over \bar M_s } = { 1 \over m} + { 1 \over M_s} \,.
\end{equation}
Thus,
\begin{eqnarray}
8 \int_0^\beta d\tau 
\int_0^\infty &dk& {1 \over k} 
 {d \over dk} \Big[ C(k,\tau) \Pi^{(0)}(k,\tau) \Big] 
\nonumber\\
&=& - 8 \, {\sum}_s e_s^2 \,
\langle n_s \rangle^{(0)}_\beta  \int_0^\beta d\tau 
\int_0^\infty dk \left\{ { 1 \over \bar M_s } \tau \Big[ 1 + \tau/\beta
\Big] \right\} 
 \exp\left\{ - { {\bf k}^2 \over 2 \bar M_s } \,
 \tau \Big[ 1 -  \tau  / \beta \, \Big] \right\}
\nonumber\\
&=& - 4  \, {\sum}_s e_s^2 \,
\langle n_s \rangle^{(0)}_\beta  \int_0^\beta d\tau \,
 \sqrt{ { 2\pi \over \bar M_s } \tau \Big[ 1 + \tau/\beta
\Big] } 
\nonumber\\
&=& - { 1 \over 8} \beta \, {\sum}_s 4 \pi e_s^2 \,
\langle n_s \rangle^{(0)}_\beta \,
\sqrt{ 2 \pi \beta \over \bar M_s } \,.
\label{clever}
\end{eqnarray}
We write
\begin{equation}
\kappa_{D,s}^2  = 4 \pi e_s^2 \langle n_s \rangle_\beta \,,
\end{equation}
which is the contribution of the plasma species $s$ to the squared
Debye wave number, and
\begin{equation}
\bar \lambda_s =  \sqrt{ 2 \pi \beta \over \bar M_s } \,,
\end{equation}
which is the thermal wave length of a particle with the reduced mass
$\bar M_s$. With these definitions, our evaluation reads
\begin{equation}
8 \int_0^\beta d\tau 
\int_0^\infty  dk {1 \over k} 
 {d \over dk} \Big[ C(k,\tau) \Pi^{(0)}(k,\tau) \Big] 
= -{ 1 \over 8} \beta \, {\sum}_s \kappa_{D,s}^2 \, \bar \lambda_s \,.
\end{equation}

Collecting the results finally yields
\begin{equation}
\langle n \rangle_\beta = e^{\beta\mu} \left( { M \over 2\pi \beta}
\right)^{3/2} \left\{ 1 + { 1 \over 2} \beta q^2 \kappa_D - { \beta
q^2 \over 16 } {\sum}_s \kappa_{D,s}^2 \, \bar \lambda_s \right\} \,.
\label{nresult}
\end{equation}
This result is accurate including terms of order $e^4$. The first
correction is the classical correction, and it is of order $e^3$. The
second correction, of order $e^4$, is of a quantum nature. It involves
the quantum wave length $\bar \lambda_s$ which, with ordinary units,
is proportional to $\hbar$.

The light electron has a thermal wave length $\lambda_e$ that is much
larger than the thermal wave lengths $\lambda_s$ of the 
nuclear particles in the plasma. Thus, the electron provides the
largest quantum correction in the result (\ref{nresult}). The large
thermal wave length of the electron, however, also implies that the
chemical potential factor $\exp\{\beta \mu_e \}$ for the electron is
much larger than those of the other plasma particles since electrical
neutrality requires that half the particles in the plasma be electrons
but their number density is proportional to $\lambda_e^{-3}$. Thus it
may well not be a good approximation to treat the electron
distribution as a dilute, Maxwell-Boltzmann distribution. On the other
hand, the small electron mass implies that wave numbers $k$ that are
very small in comparison to $\sqrt{M_s / \beta}$
dominate in the electron contribution to Eq.~(\ref{clever}), and so for
their contribution we may write
\begin{equation}
C(k,\tau) \simeq 1 \,,
\end{equation}
up to a correction of relative order $m_e/M_s$. In this case, only
the zero frequency mode of the electron polarization function appears,
and the contribution of this function can easily be evaluated since
the electron Fermi-Dirac distribution may be written as a superposition
of Maxwell-Boltzmann distributions,
\begin{equation}
{ 1 \over e^{ \beta [ E_e(p) - \mu_e ] } + 1 } =
\sum_{n=1}^\infty (-1)^{n+1} e^{ - n\beta [ E_e(p) - \mu_e ] } \,.
\end{equation}
Thus the electron contribution to Eq.~(\ref{clever}) may be expressed
as an infinite sum of the Maxwell distributions already used in the
previous evaluation, and we have
\begin{eqnarray}
8 \int_0^\beta d\tau 
\int_0^\infty  dk {1 \over k} 
 {d \over dk}  \Pi_e^{(0)}(k,\tau) 
&=& 8 \sum_{n=1}^\infty (-1)^{n+1} \int_0^{n\beta} d\tau 
\int_0^\infty  dk {1 \over k} 
 {d \over dk}  \Pi^{\rm (M-B)}(k,\tau ; n\beta ) 
\nonumber\\
&=& - { \pi e^2 \over 2} \sum_{n=1}^\infty (-1)^{n+1} 
\langle n_e \rangle^{\rm (M-B)}_{n\beta} \, n\beta 
\sqrt{ 2 \pi n\beta \over m_e } 
\nonumber\\
&=& - { 1 \over 4} e^2 m_e \sum_{n=1}^\infty (-1)^{n+1} 
2 e^{n\beta \mu_e} 
\nonumber\\
&=& - { 1 \over 4} e^2 m_e { 2 \over e^{-\beta \mu_e} + 1 } \,,
\end{eqnarray}
where the extra factor of 2 accounts for the two spin polarizations of
the electron. 
Using this more accurate treatment for the electrons and noting that
\begin{equation}
a_0 = { 1 \over e^2 m_e} 
\end{equation}
is the electron Bohr radius, we find that
\begin{equation}
\langle n \rangle_\beta = e^{\beta\mu} \left( { M \over 2\pi \beta}
\right)^{3/2} \left\{ 1 + { 1 \over 2} \beta q^2 \kappa_D - { \beta
q^2 \over 16 } \sum_{s\neq e} \kappa_{D,s}^2 \, \bar \lambda_s 
- { 1 \over 4} \beta { q^2 \over  a_0 } { 1 \over e^{ -\beta \mu_e} + 1
} \right\} \,.
\end{equation}
It should be recalled that we consider different spin states as
separate species; the total number density of a spin 1/2 species is
given by twice this result.

\section{Classical Plasma}

As discussed in the text, the correction due to the ionic component of
the plasma cannot be evaluated using a classical treatment for the
plasma unless its thermal wavelength is much less than the turning
radius of the reacting particles. In the text, we extracted this limit
from our general result for a weakly coupled plasma. Here we shall
show how this result is obtained when one starts with a formulation
that is often employed in the literature. This formulation neglects
the motion of the center of mass and takes the reaction rate to be
proportional to the single-particle, relative motion
quantum-mechanical expectation value
\begin{eqnarray}
\Bigg \langle &\,& \Bigg \langle \check {\bf r}' = 0 \Bigg | 
\exp\left\{ - \beta \left[ H_r +  e_1 \phi(\check {\bf r} M_2/M) 
+ e_2 \phi(\check {\bf r} M_1/M) \right] \right\} \Bigg| 
\check {\bf r}' = 0 \Bigg \rangle \Bigg\rangle_\beta^{\rm cl} \propto
\nonumber\\
&& \Bigg( \Bigg|
\left\langle \left( \exp\left\{ - \int_0^\beta d\tau 
\Big[ e_1 \phi(\check {\bf r}(\tau) M_2 /M) 
+ e_2 \phi(\check {\bf r}(\tau) M_1/M ) \Big]
\right\} \right)_+ \right\rangle_\beta^{\rm cl} 
\Bigg| \Bigg) \,,
\end{eqnarray}
where we have passed to the interaction picture on the right-hand side
of the equation [{\it c.f}.\ Eq.~(\ref{intpic})] and used the
normalized expectation value defined in Eq.~(\ref{normed}).  Here
classical statistical mechanics is used to evaluate the thermal
expectation value. This formula is often used in the literature to
obtain an effective potential for the quantum-mechanical relative
motion, and then the tunneling problem which describes the quantum
expectation value is solved for this effective potential. We shall
content ourselves by showing that this procedure gives our previous
limit for the first plasma correction when the plasma is dilute. Using
the result (\ref{phiclass}) for the classical thermal expectation
value of the potential correlation function and expanding out to the
first correction, we get terms of the form
\begin{eqnarray}
&& 1 + { 1 \over 2} \int_0^\beta d\tau d\tau' \left\{ e_1^2 
\Big\langle 
\phi(\check {\bf r}(\tau) M_2/M) \, \phi(\check {\bf r}(\tau') M_2/M) 
\Big\rangle_\beta^{\rm cl} + \cdots \right\} = 1 + 
\nonumber\\
&&
{ 1 \over 2} \int_0^\beta d\tau d\tau' \left\{e_1^2 { M_2/M \over \beta
| \check {\bf r}(\tau) - \check {\bf r}(\tau') | } \Big[ 1 -
\exp\{ - \kappa_D |\check {\bf r}(\tau) - \check {\bf r}(\tau')| M_2/M  
\} \, \Big] + \cdots \right\} \,.
\end{eqnarray}
Expanding in powers of the Debye wave number, which is equivalent to
expanding in the small parameter $\kappa_D r_{\rm max}$, and adding up
all the terms produces
\begin{eqnarray}
1 + {1 \over 2} (e_1 + e_2)^2 \beta \kappa_D - {\kappa_D^2 \over 4
\beta } \int_0^\beta &d\tau& d\tau' \Bigg\{ \left(
e_1^2 \,{M_2 \over M} + e_2^2 \,{M_1 \over M} \right)  
| \check {\bf r}(\tau) -\check {\bf r}(\tau') |  
\nonumber\\
&&+ 2 e_1 e_2 \left | {M_2 \over M} \check {\bf r}(\tau) +
        {M_1 \over M} \check {\bf r}(\tau') \right | \Bigg\} \,.
\label{wrong}
\end{eqnarray}
This is precisely the classical plasma limit that was discussed in the
text following Eq.~(\ref{geee}). 

\section{Real Time Troubles}

The work in the text made use of thermodynamic, ``imaginary time''
methods. Here we shall compare and contrast this method with the
formulation which employs purely real time methods. The real time
method might appear have the advantage of displaying the
dynamics of the reaction process in terms of simple physical pictures,
such as that proposed by Carrero, Sch\"afer, and Koonin (1988). We
shall show explicitly, however, that this is an illusion. Terms in the
real time formulation that apparently have a straight forward
physical interpretation may, in fact, be completely cancelled out by
other terms. In particular, we shall show that the correction found by
Carrero {\it et al.}\ has such a cancelation and hence does not exist.

To relate the two formulations, we shall pass to an interaction
picture. This is done by partitioning the complete Hamiltonian $H$ of
the total system into a part $H_0$ that describes the dynamics of the
background plasma and the reacting particles, but with no interactions
between the reacting particles and the plasma, and the remaining part
$H_1$ that the describes the interactions of the reacting particles
with the background plasma, 
\begin{equation}
H = H_0 + H_1 \,.
\end{equation}
The interaction picture is obtained by writing
\begin{equation}
e^{ -i H (t_1 -t_2)} = e^{-iH_0 t_1} U_+(t_1,t_2) e^{iH_0 t_2} \,,
\label{intpict}
\end{equation}
where
\begin{equation}
U_+(t_1,t_2) = \left( \exp\left\{ -i \int_{t_1}^{t_2} dt H_1(t) \right\}
\right)_+ 
\end{equation}
involves the time-ordered exponential with
\begin{equation}
H_1(t) = e^{ iH_0 t } H_1 e^{ -i H_0 t } \,.
\end{equation}
The interaction picture time evolution operator is unitary,
\begin{equation}
U_+(t_1,t_2)^\dagger = U_+(t_1,t_2)^{-1} \,,
\label{inverse}
\end{equation}
and obeys the group property,
\begin{equation}
U_+(t_1,t_3) U_+(t_3,t_2) = U_+(t_1,t_2) \,.
\label{group}
\end{equation}

The general expression for the reaction rate was derived in the text
and presented in Eq.~(\ref{genresult}), which we repeat here for
convenience: 
\begin{eqnarray}
\Gamma &=& \int_{-\infty}^{+\infty} dt \int (d{\bf r}) \left\langle
 {\cal K}^\dagger({\bf r},t) {\cal K}(0) 
\right\rangle_\beta 
\nonumber\\
        &=& g^2 \int_{-\infty}^{+\infty} dt \int (d{\bf r}) \, e^{iQt} 
\left\langle \psi^\dagger_1({\bf r},t) \, \psi^\dagger_2({\bf r},t) \,
\psi_3({\bf r},t) \, \psi_4({\bf r},t) \, \psi^\dagger_4(0) \,
\psi^\dagger_3(0) \, \psi_2(0) \, \psi_1(0) 
\right\rangle_\beta \,.
\nonumber\\
&& \null
\label{repeat}
\end{eqnarray}
Using the interaction picture decomposition (\ref{intpict}) for the
statistical operator $\exp\{ - \beta H \}$ which is the weight in the
thermal expectation value $\langle \cdots \rangle_\beta$ and also
using this decomposition for the time dependence of 
$ {\cal K}^\dagger({\bf r}, t) $ produces
\begin{equation}
\Gamma = \int_{-\infty}^{+\infty} dt \int (d{\bf r}) \left\langle
U_+(-i\beta , 0) U_+^{-1}(t,0) {\cal K}^\dagger({\bf r},t) U_+(t,0) 
{\cal K}(0) \right\rangle_\beta^I \,.
\label{imtime}
\end{equation}
Here the superscript $I$ indicates that the time dependence is now
governed by $H_0$ and also that the statistical ensemble is now
described (except for the full normalizing partition function in the
denominator) by $\exp\{ - \beta H_0 \}$. This is the ``imaginary
time'' formulation that was essentially employed in the text. 

In the real time formulation, one computes the thermal average of
the square of corrected matrix elements and thus arrives at
\begin{eqnarray}
\Gamma = \int_{-\infty}^{+\infty} dt \int (d{\bf r}) && \Big\langle
\Big[ U_+^\dagger(t,-\infty) {\cal K}^\dagger({\bf r},t) 
U_+^\dagger(+\infty , t) \Big] 
\nonumber\\
&& \qquad\qquad 
\Big[ U_+(+\infty , 0) {\cal K}(0) U_+(0, -\infty) \Big] 
\Big\rangle_\beta^I \,.
\label{retime}
\end{eqnarray}
To prove that this is indeed the same as the previous result
(\ref{imtime}), we make use of the group property (\ref{group})
and unitarity (\ref{inverse}) to write Eq.~(\ref{retime}) as
\begin{equation}
\Gamma = \int_{-\infty}^{+\infty} dt \int (d{\bf r})  \left\langle
 U_+^\dagger(0,-\infty) U_+(t,0)^{-1} {\cal K}^\dagger({\bf r},t) 
U_+^\dagger(t,0)  {\cal K}(0) U_+(0, -\infty) 
\right\rangle_\beta^I \,.
\label{temp}
\end{equation}
Using the cyclic symmetry of the trace which defines the thermal
average, we encounter
\begin{equation}
U_+(0, -\infty) e^{ -\beta H_0 } U_+^{-1}(0,-\infty)
= e^{ -\beta H_0 } U_+( -i\beta , -\infty -i\beta ) U_+^{-1}( 0 ,
-\infty) \,.
\end{equation}
It is implicit in the interaction picture representation that the
interaction is adiabatically damped at large times (the ``$i\epsilon$
prescription''). Thus we may replace the complex infinite time limit $
-\infty -i \beta $ by the real limit $ - \infty$ and use the group
property (\ref{group}) to conclude that
\begin{equation}
U_+(0, -\infty) e^{ -\beta H_0 } U_+^{-1}(0,-\infty)
= e^{ -\beta H_0 } U_+( -i\beta , 0 ) \,.
\end{equation}
Using these results in Eq.~(\ref{temp}) reduces it to the imaginary
time form (\ref{imtime}) and proves its equivalence with the real time
form (\ref{retime}).

To illustrate the differences in the real and imaginary time
formulations, we shall examine a simple model in some detail using the
methods of both forms. In this model, all the interactions of the
final, produced particles are neglected (which may be done if the
energy release $Q$ is large). Thus, in this model,
\begin{equation}
H_1 = \int ( d{\bf r}) [ \rho_1 \phi + \rho_2 \phi ] \,,
\end{equation}
where $\rho_1$ and $\rho_2$ are the charge density operators of the
initial particles labeled by 1 and 2. To simplify the exposition, we
shall also assume that one of the initial particles, say 2, is very
massive. Thus, except for its ordering position, the charge density
operator $\rho_2$ is independent of time. To further simplify the
model, we shall neglect the Coulomb interactions between the initial
particles 1,2 and compute only the correction involving a single
interaction between each initial particle and the background plasma,
the term proportional to $e_1e_2$ which we denote by $\Gamma_{12}$. 
Finally, in our model we shall take the limit in which the background
plasma is treated classically. In this classical limit, the wave number
vector ${\bf k}$ of a plasma correlation function is generally neglected
relative to an initial particle momentum ${\bf p}$ since these enter
in the combination $ {\bf p} + \hbar {\bf k}$. As we shall see, this
neglect of the plasma wave number may be done and gives the classical
limit except when energy corrections are involved which lead to
frequency shifts depending upon $\Delta E/\hbar$.

We first compute the $e_1 e_2$ term in the imaginary time formulation
(\ref{imtime}).  As discussed in the text, the corrections associated
with the time evolution operator $U_+(t,0)$ are negligible. We neglect
these here and compute only those associated with the statistical
factor $U_+(-i\beta,0)$, 
\begin{eqnarray}
\Gamma_{12} &=&  \int_{-\infty}^{+\infty} dt \int (d{\bf r}) 
\int_0^\beta d\tau_1 d\tau_2 \int (d{\bf r}_1) (d{\bf r}_2) 
\left\langle \Big( \phi({\bf r}_1,\tau_1) \phi({\bf r}_2,\tau_2) \Big)_+
\right\rangle_\beta 
\nonumber\\
&& \qquad\qquad
\left\langle \Big( \rho_1({\bf r}_1,\tau_1) \rho_2({\bf r}_2,\tau_2) 
\Big)_+  {\cal K}^\dagger({\bf r},t) {\cal K}(0) 
\right\rangle_\beta^I \,.
\label{im12}
\end{eqnarray}
Since the interaction between particles 1 and 2 is neglected, the
charge density operators $\rho_1$ and $\rho_2$ commute. Hence, since
$\rho_2$ is time independent, the time-ordered product of the two
charge density operators is independent of $\tau_2$. In view of the
discussion of Appendix A, the correction (\ref{im12}) thus involves
\begin{eqnarray}
\int_0^\beta  d\tau_2 
\left\langle \Big( \phi({\bf r}_1,\tau_1) \phi({\bf r}_2,\tau_2) \Big)_+
\right\rangle_\beta 
&=& \int { (d{\bf k}) \over (2\pi)^3 } { 4\pi \over k^2 } \left[ 1 - { 1
\over \epsilon(k,0) } \right] e^{ i{\bf k} \cdot ( {\bf r}_1 - {\bf
r}_2 ) } 
\nonumber\\
&=& \beta \left\langle  \phi({\bf r}_1) \phi({\bf r}_2) 
\right\rangle_\beta^{\rm cl} \,, 
\end{eqnarray}
where in the second line we have written the classical limit. 
We stress that, without any further approximation, this correction
entails the dielectric permittivity of the plasma $\epsilon(k,\omega)$
at zero frequency, $\omega =0$. That is, the correction depends only
upon the {\it static} properties of the plasma. Now, in the classical
limit, the plasma wave number ${\bf k}$ is neglected, and the
correction entails only the total charge operators of the initial
particles,
\begin{equation}
\int (d{\bf r}_1) \rho_1({\bf r}_1,\tau) = Q_1 \,, \qquad
\int (d{\bf r}_2) \rho_2({\bf r}_2,\tau) = Q_2 \,,
\end{equation}
which are time independent. 
In the thermal expectation value in Eq.~(\ref{im12}), these operators
just measure the charge (with the Boltzmann statistics for the dilute
initial particles which we always take), and so we may make the
replacements $Q_1 \to e_1$ and $Q_2 \to e_2$ to obtain
\begin{equation}
\Gamma_{12} = e_1 e_2 \beta^2 \left\langle  \phi({\bf 0}) \phi({\bf 0}) 
\right\rangle_\beta^{\rm cl} \Gamma_0 \,,
\end{equation}
where $\Gamma_0$ is the free gas reaction rate. The classical
correlation function for a dilute plasma gives, according to
Eq.~(\ref{phiclass}), 
\begin{equation}
\beta \left\langle  \phi({\bf 0}) \phi({\bf 0}) 
\right\rangle_\beta^{\rm cl} = \kappa_D \,,
\end{equation}
and thus we arrive at the basic Salpeter correction,
\begin{equation}
\Gamma_{12} = e_1 e_2 \beta \kappa_D \Gamma_0 \,.
\end{equation}
Let us note for future reference that, as discussed in the
Introduction,  
\begin{equation}
U_{12}^{{\rm static \,\, pol}} = - e_1 e_2 \kappa_D 
\end{equation}
is just the {\it static} polarization energy of the plasma induced
when the two initial particles are on top of one another. Thus the
Salpeter result may be written as 
\begin{equation}
\Gamma_{12} = - \beta U_{12}^{{\rm static \,\, pol}} \Gamma_0 \,.
\label{static}
\end{equation}

We turn now to compute the $e_1 e_2$ term from the real time form
(\ref{retime}), which gives, on reading from right to left,
\begin{eqnarray}
\Gamma_{12} &=&  \int_{-\infty}^{+\infty} dt \int (d{\bf r}) 
\int_{-\infty}^{+\infty} dt_1 dt_2 \int (d{\bf r}_1) (d{\bf r}_2) 
\nonumber\\
&& 
\Bigg\{ - \Big\langle 
\Big( \phi({\bf r}_1,t_1) \phi({\bf r}_2,t_2) \Big)_+
\Big\rangle_\beta 
\left\langle   {\cal K}^\dagger({\bf r},t) 
\Big( \rho_1({\bf r}_1,t_1) \rho_2({\bf r}_2,t_2) 
 {\cal K}(0) \Big)_+
\right\rangle_\beta^I 
\nonumber\\
&& 
+ \Big\langle 
\phi({\bf r}_2,t_2)  \phi({\bf r}_1,t_1) 
\Big\rangle_\beta 
\left\langle \Big( \rho_2({\bf r}_2,t_2) 
 {\cal K}^\dagger({\bf r},t) \Big)_- \Big( \rho_1({\bf r}_1,t_1) 
{\cal K}(0) \Big)_+ 
\right\rangle_\beta^I 
\nonumber\\
&& 
+ \Big\langle 
 \phi({\bf r}_1,t_1) \phi({\bf r}_2,t_2) 
\Big\rangle_\beta 
\left\langle \Big( \rho_1({\bf r}_1,t_1) 
 {\cal K}^\dagger({\bf r},t) \Big)_- \Big( \rho_2({\bf r}_2,t_2) 
{\cal K}(0) \Big)_+ 
\right\rangle_\beta^I 
\nonumber\\
&& 
 - \Big\langle 
\Big( \phi({\bf r}_1,t_1) \phi({\bf r}_2,t_2) \Big)_-
\Big\rangle_\beta 
\left\langle \Big( \rho_1({\bf r}_1,t_1) \rho_2({\bf r}_2,t_2) 
  {\cal K}^\dagger({\bf r},t) \Big)_- {\cal K}(0) 
\right\rangle_\beta^I \Bigg\} \,,
\label{messy}
\end{eqnarray}
where the notation $( \cdots )_-$ denotes the anti-time-ordered
product.  To simplify the work, we immediately take the classical
limit in which the momentum transfer $\hbar {\bf k}$ imparted by the
plasma to the heavy particle charge density $\rho_2$ vanishes. In this
limit, the integration over ${\bf r}_2$ in Eq.~(\ref{messy}) gives the
total charge $Q_2(t_2)$ which is time independent save for its
ordering.  With the Boltzmann statistics that we use for the reacting
particles 1,2, $Q_2$ must appear to the left of the creation operator
$\psi_2^\dagger$ and to the right of the destruction operator
$\psi_2$, and in these cases it is simply replaced by $e_2$.
This restriction of the ordering of $Q_2(t_2)$ limits the
time interval over which $t_2$ runs to $ -\infty < t_2 < 0$ or $
-\infty < t_2 < t$. The interval from $0$ to $t$ described by the
time evolution operator $U_+(t,0)$ will be neglected here just has it
has been previously.  With the Boltzmann statistics, a similar time 
restriction obtains for $\rho({\bf r}_1,t_1)$, $ -\infty < t_1 < 0$. 
Thus, the real time result (\ref{messy}) reduces to
\begin{eqnarray}
\Gamma_{12} &=&  \int_{-\infty}^{+\infty} dt \int (d{\bf r}) 
\int_{-\infty}^0 dt_1 \int_{-\infty}^0 dt_2 \int (d{\bf r}_1) 
\nonumber\\
&& \qquad \Bigg\{ 
- \Big\langle 
\Big( \phi({\bf r}_1,t_1) \phi({\bf 0},t_2) \Big)_+
\Big\rangle_\beta 
e_2 \left\langle   {\cal K}^\dagger({\bf r},t) 
 {\cal K}(0)  \rho_1({\bf r}_1,t_1) 
\right\rangle_\beta^I 
\nonumber\\
&& \qquad\quad
+ \Big\langle 
\phi({\bf 0},t_2)  \phi({\bf r}_1,t_1) 
\Big\rangle_\beta 
e_2 \left\langle 
 {\cal K}^\dagger({\bf r},t) 
{\cal K}(0)  \rho_1({\bf r}_1,t_1) 
\right\rangle_\beta^I 
\nonumber\\
&& \quad\qquad
+ \Big\langle 
 \phi({\bf r}_1,t_1) \phi({\bf 0},t_2) 
\Big\rangle_\beta 
e_2 \left\langle  \rho_1({\bf r}_1,t_1) 
 {\cal K}^\dagger({\bf r},t) 
{\cal K}(0) 
\right\rangle_\beta^I 
\nonumber\\
&& \quad\qquad
 - \Big\langle 
\Big( \phi({\bf r}_1,t_1) \phi({\bf 0},t_2) \Big)_-
\Big\rangle_\beta 
e_2 \left\langle  \rho_1({\bf r}_1,t_1) 
  {\cal K}^\dagger({\bf r},t)  {\cal K}(0) 
\right\rangle_\beta^I \Bigg\} \,.
\label{lessmess}
\end{eqnarray}
The terms that appear in the successive lines on the right-hand side
of this equation appear to have a clear physical description. The
first line corresponds to the correction that results when the initial
particles 1 and 2 have a Coulomb exchange which is modified by the
plasma in a dynamical (frequency or time dependent) fashion before the
fusion reaction takes place. The last line give the same correction
for the complex conjugate amplitude that enters into the squared
matrix element. The second and third lines do not describe such
second-order interactions which take place between the initial
particles before the fusion occurs. They instead describe the square
of first-order amplitudes. They correspond to processes in which, if a
complete set of final plasma states is introduced between ${\cal
K^\dagger}$ and ${\cal K}$, the plasma is left in excited states. In
one of the amplitudes, this excitation is caused by particle 1, in
the other, by particle 2. The trouble with the real-time formulation
is that there is extensive cancelation between these terms that might
appear to have a distinct physical meaning.

To make a first combination of terms, we write
\begin{equation}
\left\langle \Big( \phi({\bf r},t) \phi({\bf r}',t') \Big)_\pm 
\right\rangle_\beta = \pm {1\over2} \epsilon(t-t')
\left\langle \Big[ \phi({\bf r},t) ,\phi({\bf r}',t') \Big] 
\right\rangle_\beta + {1\over2} 
\left\langle \Big\{ \phi({\bf r},t) ,\phi({\bf r}',t') \Big\}
\right\rangle_\beta \,,
\label{tord}
\end{equation}
where the curly brackets denote the anticommutator, and 
\begin{equation}
\Big\langle  \phi({\bf r},t) \phi({\bf r}',t') 
\Big\rangle_\beta = {1\over2} 
\left\langle \Big[ \phi({\bf r},t) ,\phi({\bf r}',t') \Big] 
\right\rangle_\beta + {1\over2} 
\left\langle \Big\{ \phi({\bf r},t) ,\phi({\bf r}',t') \Big\}
\right\rangle_\beta \,.
\end{equation}
The terms involving the anticommutator pieces in Eq.~(\ref{lessmess})
all cancel. As discussed in Appendix A, the commutator has the thermal
expectation value
\begin{equation}
\left\langle \Big[ \phi({\bf r},t) ,\phi({\bf r}',t') \Big]
\right\rangle_\beta =
\int {(d{\bf k}) \over (2\pi)^3} \int {d\omega' \over 2\pi} \left(
{4\pi \over k^2} \right)^2 c(k,\omega') e^{i{\bf k} \cdot ({\bf r} -
{\bf r}') - i\omega' (t-t')} \,.
\end{equation}
Appendix A also shows that $c(k,\omega')$ appears as the spectral
weight which defines the dielectric function $\epsilon(k,\omega)$ by
the dispersion relation
\begin{equation}
{1 \over \epsilon(k,\omega)} -1 = {4\pi \over k^2} \int {d\omega'
\over 2\pi} {c(k,\omega') \over \omega - \omega' - i\epsilon} \,,
\label{dispp}
\end{equation}
a result that we shall soon make use of. 

Making use of Eq's.~(\ref{freer}) and (\ref{freel}) in the Boltzmann
statistics limit in which $n({\bf p})$ is neglected relative to $1$, and
neglecting the plasma wave number ${\bf k}$ in spatial momentum terms,
but not in the energies as is appropriate to the limit in which the
plasma in treated classically, the first two lines of the right-hand
side of Eq.(\ref{lessmess}) involve
\begin{equation}
\int (d{\bf r}_1) e^{i{\bf k} \cdot {\bf r}_1} \left\langle
\psi_1^\dagger({\bf r},t) \psi_1(0) \rho_1({\bf r}_1,t_1)
\right\rangle_\beta = e_1 \int {(d{\bf p}) \over (2\pi)^3} n_1({\bf p})
e^{-i{\bf p} \cdot {\bf r} +iE_1({\bf p}) t} e^{i\Delta E t_1} \,,
\end{equation}
and the second two lines entail
\begin{equation}
\int (d{\bf r}_1) e^{i{\bf k} \cdot {\bf r}_1} \left\langle
\rho_1({\bf r}_1,t_1) \psi_1^\dagger({\bf r},t) \psi_1(0) 
\right\rangle_\beta = e_1 \int {(d{\bf p}) \over (2\pi)^3} n_1({\bf p}
+ {\bf k})
e^{-i{\bf p} \cdot {\bf r} +iE_1({\bf p}) t} e^{i\Delta E t_1} \,,
\end{equation}
in which 
\begin{eqnarray}
\Delta E &=& E_1({\bf p} + {\bf k}) - E_1({\bf p}) 
= { 2{\bf p} \cdot {\bf k} + {\bf k}^2 \over 2m_1} 
\nonumber\\
&\simeq& 
{ {\bf p}
\cdot {\bf k} \over m_1} = {\bf v}_1 \cdot {\bf k} \,,
\end{eqnarray}
where the approximate equality applies to our classical limit. 
In view of these results and the decomposition (\ref{tord}), we see
that the first and last lines of Eq.~(\ref{lessmess}) contain the
combination
\begin{equation}
- n_1({\bf p}) + n_1({\bf p} + {\bf k}) = - e^{-\beta E({\bf p})} +
e^{-\beta E({\bf p} + {\bf k})} \simeq
- \beta \Delta E \, n_1({\bf p}) \,,
\end{equation}
where, as always, we use Boltzmann statistics for the initial
reacting particles, and the last approximation is appropriate for our
classical plasma limit. 
The first and last lines on the right-hand side of
Eq.~(\ref{lessmess}) are evaluated with the time integrals
\begin{equation}
\int_{-\infty}^0 dt_1 \int_{-\infty}^0 dt_2 \epsilon(t_1-t_2)
e^{-i\omega'(t_1-t_2)} e^{i\Delta E t_1} = {1 \over \Delta E} \left[
{1 \over \Delta E - \omega' -i\epsilon} + {1 \over -\omega'} \right] \,.
\label{oddt}
\end{equation}
The integration of the denominators which appear here over $\omega'$ with
the weight $c(k,\omega')$ is just the dispersion relation
(\ref{dispp}), and we see that the first and last lines of
Eq.~(\ref{lessmess}) produce $\beta$ times the factor
\begin{equation}
- U_{12}^{\rm dyn \,\, pol}({\bf v}_1) =  - {1 \over 2} e_1 e_2
\int {(d{\bf k}) \over (2\pi)^3} {4\pi \over k^2} \left\{ \left[ {1
\over \epsilon(k,{\bf v}_1 \cdot {\bf k})} - 1 \right] +
\left[ {1 \over \epsilon(k,0)} - 1 \right] \right\} \,.
\label{almost}
\end{equation}
We schematically indicate the resulting correction to the reaction
rate in terms of an effective average value of this dynamical plasma
polarization by writing
\begin{equation}
\Gamma_{12}^{\rm dyn \,\, pol} = - \beta \left\langle U_{12}^{\rm dyn
\,\, pol}({\bf v}_1) \right\rangle \Gamma_0 \,.
\end{equation}

This is essentially the result obtained by Carrero, Sch\"afer, and
Koonin (1988) (to order $e_1 e_2$). It corresponds to using the $e_1
e_2$ term of the classical, 
dynamical plasma polarization energy
\begin{equation}
U(t) = {1\over2} \int (d{\bf r}) \rho({\bf r},t) \phi^{\rm pol}({\bf r},t)
\end{equation}
evaluated for
\begin{equation}
\rho({\bf r},t) = e_1 \delta( {\bf r} - {\bf v}_1 t) +e_2 \delta({\bf
r}) \,,
\end{equation}
at $t=0$ when the two initial particles are on top of one another,
\begin{equation}
U_{12}^{\rm dyn \,\, pol}({\bf v}_1) = e_2 \phi^{\rm pol}_1({\bf 0},0) +
e_1 \phi^{\rm pol}_2({\bf 0},0) \,.
\end{equation}

This result, however, is not the whole story. The dynamical aspects of
this result, which corresponds to the second-order Coulomb exchange
between the two initial reacting particles as modified by the plasma,
are completely cancelled by the second and third lines of
Eq.~(\ref{lessmess}), which correspond to squares of first-order
amplitudes that leave the plasma in excited states. These terms have
exactly the same structure as the dynamical terms that have just been
computed except that the time integration (\ref{oddt}) is now replaced
by
\begin{equation}
\int_{-\infty}^0 dt_1 \int_{-\infty}^0 dt_2 
e^{-i\omega'(t_1-t_2)} e^{i\Delta E t_1} = {1 \over \Delta E} \left[ - 
{1 \over \Delta E - \omega' -i\epsilon} + {1 \over -\omega'} \right] \,.
\end{equation}
This gives rise to a correction which is the same form as that in
Eq.~(\ref{almost}) except that the terms in the curly braces there are
replaced by 
\begin{equation}
\left\{ - \left[ {1
\over \epsilon(k,{\bf v}_1 \cdot {\bf k})} - 1 \right] +
\left[ {1 \over \epsilon(k,0)} - 1 \right] \right\} \,.
\label{done}
\end{equation}
The addition of these terms cancels the dependence upon the
frequency-dependent $\epsilon(k,{\bf v}_1 \cdot {\bf k})$ and yields
precisely the Salpeter result (\ref{static}). The moral to the story
is that simple physical pictures should be augmented with a correct
formal basis to ensure that correct physical results are obtained.  In
addition to explaining the discrepancy between our results and those
of Carrero {\it et al}.\ (1988), this cancelation may serve as a
general caveat with respect to approaches that formulate the problem
completely in terms of a plasma-modified two body interaction between
the fusing particles. We should note, however, that, in one
circumstance, the plasma excitation terms which give rise to
correction involving Eq.~(\ref{done}) vanish. The frequency --
$\omega$ -- dependence of $\epsilon(k,\omega)$ for a classical plasma
arises from a plasma particle's velocity $ {\bf v}_s $ in the form $
{\bf v}_s \cdot {\bf k}$. Thus, if the typical velocity of the
reacting particle is much less than the typical velocity of the plasma
particle, $|{\bf v}_1| \ll |{\bf v}_s|$, then $\epsilon(k,{\bf v}_1
\cdot {\bf k})$ may be replaced by $\epsilon(k,0)$, and the plasma
excitation correction involving Eq.~(\ref{done}) vanishes. This is the
case when the mass of the plasma particle $M_s$ is much less than that
of the reacting particle, $M_s \ll M_1$, as are the electrons in the
plasma.

\section{Dynamical Correction Size}

Here we shall obtain the order of magnitude of the leading
``dynamical'' correction given in Eq.~(\ref{correctt}). To do this, we
take the dilute plasma limit to evaluate the real-time correlation
function that appears in Eq.~(\ref{correctt}) as
\begin{equation}
\langle \phi(0) \phi( {\bf r},t') \rangle_\beta
= \int { (d {\bf k}) \over (2\pi)^3 } e^{- i {\bf k} \cdot {\bf r} }
\left[ { 4\pi \over k^2} \right]^2 \Pi^{(0)} (k, -t') \,,
\label{fifi}
\end{equation}
where $ \Pi^{(0)}(k, -t') $ is the plasma polarization function given
at the end of Appendix A in Eq's.~(\ref{mist}) and (\ref{pitwo}),
but continued to real time. With both $\tau - \tau' > 0$ and $t - t' >
0$ [as is appropriate to the operator ordering in Eq.~(\ref{fifi})],
this continuation is given by $\tau - \tau \to i(t-t')$, and we have
\begin{equation}
\Pi^{(0)}(k, -t') \simeq {\sum}_s e_s^2 \,
\langle n_s \rangle^{(0)}_\beta C_s(k,-t') \,,
\end{equation}
in which
\begin{equation}
C_s(k, -t') = \exp\left\{ i { k^2 \over 2M_s } t' \right\} \,,
\end{equation}
where we have neglected the correction in the exponential involving
$t'/\beta$ since in our application this is of the negligible order $
1 / \beta Q$. The wave number integration over ${\bf k}$ in
Eq.~(\ref{fifi}) may be performed by using the representation
\begin{equation}
{ 1 \over k^2 } = \int_0^\infty sds \, e^{- s k^2 } \,,
\end{equation}
interchanging integrals, and completing the square to obtain
\begin{equation}
\langle \phi(0) \phi( {\bf r},t') \rangle_\beta
=  {\sum}_s e_s^2 \, \langle n_s \rangle^{(0)}_\beta \sqrt{4 \pi} 
 \, \int_0^\infty sds \left( s - { it' \over 2M_s} \right)^{- 3/2} 
\exp\left\{ - { {\bf r}^2 \over 4 ( s - it' / 2M_s )
     } \right\} \,.
\end{equation}

The correction appearing in Eq.~(\ref{correctt}) involves the
difference of this expression for $ {\bf r} \ne 0$ and $ {\bf r} = 0
$. To estimate the size of this correction, it suffices to use the
classical approximation for the operator $ {\bf r}(t') \simeq {\bf p}
t' / M $.  Noting that the produced nuclei move much faster that the
initial fusing nuclei, we see that the largest effect of this motion
is revealed by using  $ p \sim \sqrt{ 2 M Q} $ as described in
Eq.~(\ref{pest}), with now $M$ a typical value of the mass of a
produced nucleus. We change variables with $ s = u t' / 2M_s$ to
exhibit the correction as
\begin{eqnarray}
&&\qquad\quad
\pm i \beta ( e_1 + e_2) e_r \int_0^t \langle \phi(0) 
\left[ \phi( { \bf r}(t'), t') - \phi(0) \right]
\rangle_\beta 
\nonumber\\
&& \qquad\qquad
\sim
\pm i \beta (e_1 + e_2) e_r {\sum}_s e_s^2 \, \langle n_s \rangle^{(0)}_\beta 
\sqrt{4\pi} \int_0^t dt' \sqrt{ t' \over 2 M_s } 
\nonumber\\
&& \qquad\qquad\quad
\int_0^\infty udu \, 
\left\{ 
( u - i )^{-3/2} \exp\left\{ - { M_s \over M}
{ Q t' \over u - i } \right\} - u^{-3/2} \right\} \,.
\label{finishedd}
\end{eqnarray}
The $u$ integration is well defined, and the exponent in it involves
$Qt'$ which at most is of order unity.\footnote{The factor which
  replaces $Qt'$ for the initial particles is much smaller in view of
  Eq.~(\ref{pbarr}) and the discussion below it.} The overall factor
of $ 1 / \sqrt{2 M_s} $ appears to imply that the light electrons
in the plasma give the largest contribution. However, for the
electrons, the mass ratio $ m_e / M$ in the exponent in the $u$
integral is very small, and to leading order in this ratio, the
exponential can be replaced by unity. Thus, the leading contribution
from the electrons in the plasma is independent of the nature of the
particle involved in the reaction, except for the overall factor of
the particle's charge $e_r$. Since the $\pm$ sign in
Eq.~(\ref{finishedd}) is positive for the initial fusing particles
$1,2$ [from the $W(t)$] and negative for the final produced particles
$3,4$ [from the $V(t)$], and since charge is conserved, $ e_1 + e_2 -
e_3 -e_4 = 0$, we find that the leading electron contribution
vanishes. For the remaining contributions of the ions in the plasma,
it suffices to replace the $u$ integral in Eq.~(\ref{finishedd}) by a
(complex) constant of order unity.  Since $ t \sim 1/Q$, the remaining
$t$ integral is of order
\begin{equation}
\int_0^t dt' \, \sqrt{t'} \sim Q^{- 3/2} \,.
\end{equation} 
Remembering that the Debye wave number has the form $
\kappa^2_D = 4 \pi e^2 \beta \langle n \rangle_\beta$, and that the
plasma frequency of the ions $\omega_p$ has a size given by $
\omega^2_p \sim 
\kappa_D^2 / \beta M $, we find that the leading correction is
in fact of the order
\begin{equation}
i \beta ( e_1 + e_2) e_r \int_0^t \langle \phi(0) 
\left[ \phi( { \bf r}(t'), t') - \phi(0) \right]
\rangle_\beta 
\sim \beta e_1 e_2 \kappa_D \left\{ { \omega_p \over Q } \sqrt{ 1
    \over \beta Q}  \right\} \,.
\end{equation}
Here we have neglected the niceties of mass and charge ratio factors,
denoting all of the nuclear masses by the typical mass $M$, and
replacing $(e_1 + e_2) e_r$ by $e_1 e_2$.   

\newpage

\begin{center}
REFERENCES
\end{center}

\begin{itemize}

\item[]{Alastuey, A. and B.\ Jancovici, 1978, Astrophys.\ J.\ {\bf
226}, 1034.}

\item[]{Bahcall, J.~N., 1995, {\it Solar neutrinos: What we have learned},
Physical Processes in Astrophysics, Proceedings of a Meeting in Honour
of Evry Schatzman, Edited by I.~W. Roxburgh, J.~-L. Masnou
(Springer-Verlag Berlin, Germany), pp.\ 19-36.}

\item[]{Bahcall, J.~N., 1989, {\it Neutrino Astrophysics} (Cambridge
Univ.\ Press, Cambridge, England).}

\item[]Brown, L.~S., 1992, {\it Quantum Field Theory} (Cambridge
Univ.\ Press, Cambridge, England).

\item[]Carrero, C., A.\ Sch\"afer, and S.E. Koonin, 1988, Astrophys.\
J.\ {\bf 331}, 565.

\item[]Clayton, D.~D., 1968, {\it Principles of Stellar Evolution and 
Nucleosynthesis} (McGraw-Hill, New York).

\item[]Dar, A. and G. Shaviv, 1996, Nucl. Phys. B, 
Proceedings Supplements {\bf 48}, 335.

\item[]Dewitt, H.~E., H.~C. Graboske, and M.~C. Cooper, 1973,
Astrophys.\ J.\ {\bf 181}, 457.

\item[]Fetter, A.~L. and J.~D. Walecka, 1971, {\it Quantum Theory of
Many-Particle Systems}, (McGraw-Hill, New York).

\item[]Gamow, G., 1928, Zeitschr.\ Phys.\ {\bf 51}, 204.

\item[]Graboske, H.~C., H.~E. Dewitt, A.~S. Grossman, and
M.~S. Cooper, 1973, Astrophys.\ J.\ {\bf 181}, 457.

\item[]Hata, N. and P. Langacker, 1995, Phys.\ Rev.\ D {\bf 52}, 420.

\item[]Ichimaru, S., 1993, Rev.\ Mod.\ Phys.\ {\bf 65}, 255.

\item[]Jancovici, B., 1977, Journ.\ Stat.\ Phys.\ {\bf 17}, 357.

\item[]Johnson, C.~W., E. Kolbe, S.~E. Koonin, and K. Langanke, 1992,
Astrophys.\ J.\ {\bf 392}, 320.

\item[]Salpeter, E.~E., 1954, Aust.\ J.\ Phys.\ {\bf 7}, 373.

\item[]Salpeter, E.~E. and H.~M Van Horne, 1969, Astrophys.\ J.\ {\bf
155}, 183.

\item[]Shaviv, N.\ J.\ and G.\ Shaviv, 1996, Astrophys.\ J.\ {\bf 468},
433.

\end{itemize}

\end{document}